\documentclass[10pt]{article}

\usepackage[utf8]{inputenc}
\usepackage[english]{babel}
\usepackage[T1]{fontenc}
\usepackage{lmodern}
\usepackage[babel=true]{microtype}

\usepackage{amsmath,amsfonts,amssymb,amsthm}

\usepackage{graphicx}
\usepackage{color}
\usepackage[usenames,dvipsnames,svgnames,table]{xcolor}
\usepackage{float}
\usepackage{hyperref}
\usepackage{dsfont}
\usepackage{subcaption}
\usepackage[format=plain,font=small,textfont=it]{caption}
\usepackage{color}
\usepackage{multirow} 
\usepackage{bookmark}
\usepackage{mathrsfs}
\usepackage{physics}
\usepackage{multirow}
\usepackage[inline]{enumitem}
\usepackage{framed}

\usepackage[usecourier=false,theme=jupyter,charsperline=100]{jlcode}

\usepackage{parskip}

\usepackage{geometry}
\newcommand{\Wfour}[9]
{
    \left(\begin{array}{cccc} #1 & #2 & #3 & #4 \\ #5 & #6 & #7 & #8 \end{array}\right)^{(#9)}
}

\newcommand{\Wsix}[6]
{
    \left \{ \begin{array}{ccc} #1 & #2 & #3 \\ #4 & #5 & #6 \end{array}\right \}
}

\newcommand{\nn}{\nonumber}

\newcommand{\SU}{\mathsf{SU}(2)}
\newcommand{\SL}{\mathsf{SL(2, \mathbb{C})}}

\newcommand{\I}{{i\mkern1mu}}
\newcommand{\dtre}{{\Delta_3}}

\newcommand{\lib}{\texttt{sl2cfoam-next}}
\newcommand{\codek}[1]{\texttt{#1}}

\usepackage[style=numeric,
            sorting=none,
            url=false,
            doi=false,
            isbn=false]{biblatex}
\graphicspath{{img/}}

\begin{document}

\title{A high-performance code for EPRL spin foam amplitudes}

\author{\Large{Francesco Gozzini\footnote{gozzini@cpt.univ-mrs.fr} }
\smallskip \\ 
\small{CPT, Aix-Marseille\,Universit\'e, Universit\'e\,de\,Toulon, CNRS, 13288 Marseille, France}
}

\date{}

\maketitle

\begin{abstract}
\noindent We present \codek{sl2cfoam-next}, a high-performance software library for computing Lorentzian EPRL spin foam amplitudes. The library improves on previous codes by many orders of magnitude in single-core performance, can be parallelized on a large number of CPUs and on the GPU, and can be used interactively. We describe the techniques used in the code and provide many usage examples. As first applications, we use \codek{sl2cfoam-next} to complete the numerical test of the Lorentzian single-vertex asymptotics and to confirm the presence of the ``flatness problem'' of spin foam models in the BF and EPRL cases.
\end{abstract}
    






\section{Introduction}
\label{sec:intro}

Spin foam models are a tentative regularization of the Feynman path integral for the gravitational field using a background-independent discretization \cite{baezIntroductionSpinFoam2000}. The main ingredients of such theories are the formulation of General Relativity as a topological theory with constraints and the implementation of the constraints on a cellular decomposition of the spacetime manifold. The quantum theory then follows from the discrete path integral while the continuum theory must be recovered in the double limit of finer discretization and vanishing $\hbar$.

The EPRL-FK model \cite{engleLQGVertexFinite2008,freidelNewSpinFoam2008} (EPRL in the following for brevity) is the most promising spin foam model so far. Its properties have been studied mainly in the large spin regime and in simple configurations comprising one to few vertices \cite{barrettAsymptoticAnalysisEnglePereiraRovelliLivine2009,magliaroCurvatureSpinfoams2011a,rielloSelfenergyLorentzianEnglePereiraRovelliLivine2013}. Among the known results, the emergence of the Regge action \cite{barrettAsymptoticAnalysisEnglePereiraRovelliLivine2009,barrettLorentzianSpinFoam2010} and the recovery of the graviton propagator of Regge calculus \cite{bianchiLQGPropagatorNew2009} in the semiclassical limit are considered the most relevant and promising for connecting the model to the classical theory.

The analytical study of Lorentzian EPRL amplitudes is hard. The proliferation of spin and intertwiner labels, the difficulty of dealing with non-compact gauge groups and the difficulties with working outside of the saddle-point approximation of the semiclassical regime are all factors that limit the understanding of anything but the simplest configurations. This is an unfortunate situation since it is expected that non-trivial configurations are needed to study the dynamics of the theory or to model possible quantum gravitational phenomena \cite{bianchiSpinfoamCosmology2010, christodoulouPlanckStarTunneling2016,dambrosioEndBlackHole2020,soltaniEndBlackHole2021}.

Recently, the problem of dealing with Lorentzian EPRL amplitudes started to be approached from a numerical standpoint. Montecarlo integration on Lefschetz thimbles has been introduced in \cite{hanSpinfoamLefschetzThimble2021} for studying the 2-point propagator in the regime of large spins. Complementary, following the formulation of the EPRL model in \cite{spezialeBoostingWignerNjsymbols2017} the software library \codek{sl2cfoam} \cite{donaNumericalMethodsEPRL2018} has been developed for computing EPRL amplitudes with a generic number of vertices and boundary data. The first results have been very encouraging, with applications to the single-vertex Euclidean and Lorentzian asymptotics \cite{donaSUGraphInvariants2018, donaNumericalStudyLorentzian2019}, the estimation of infrared divergences \cite{donaInfraredDivergencesEPRLFK2018}, the simulation of a simple model of spin foam cosmology \cite{gozziniPrimordialFluctuationsQuantum2021} and the study of a configuration with 3 vertices \cite{donaNumericalAnalysisSpin2020,donaSearchingClassicalGeometries2020} (although using the vertex the topological BF theory). However, all the cited examples, while clearly showing the potential and advantages of dealing numerically with the complexities of the EPRL model, also highlighted that more powerful codes are needed to meet the challenges that come with more vertices or larger spin labels. In particular, the simulation of the Lorentzian sector of the asymptotic regime \cite{donaNumericalStudyLorentzian2019} and the computation of the three-vertices configuration with ``large'' spins \cite{donaNumericalAnalysisSpin2020} have been completed only partially, and the results so far have been encouraging but inconclusive. 

In this work we present \codek{sl2cfoam-next}, a completely rewritten version of the previous library \codek{sl2cfoam}. The new code uses the same formulation of the EPRL model \cite{spezialeBoostingWignerNjsymbols2017} but combines it with ideas and techniques borrowed from the field of High Performance Computing to realize a major performance improvement. The new code is faster, more robust and more user-friendly, with the possibility of computing EPRL amplitudes interactively using the Julia scripting language. As first and immediate applications, we use \codek{sl2cfoam-next} to complete the study of the Lorentzian asymptotics and of the ``flatness problem'' \cite{conradySemiclassicalLimit4d2008, bonzomSpinFoamModels2009,hellmannHolonomySpinFoam2013a, hanSpinfoamModelsLarge2014} for the spin foam with three-vertices. We stress that these are relatively simple applications for the new code\footnote{Of course, we do not consider here the complex but necessary process of understanding and modeling the two problems, which has been completed in the previous works. We just refer to the implementation in software.}, requiring only a modest investment of time and resources.

The paper is organized as follows. In Section \ref{sec:eprl} we briefly review the definition of the EPRL model and its splitting as a convergent sum of products of booster functions and $15j$-symbols computed over virtual spin labels. In Section \ref{sec:library} we provide a technical introduction to the new code along with some simple examples, benchmarks and comparisons with the previous code. Section \ref{sec:applications} contains the first applications of \codek{sl2cfoam-next} to two open problems from spin foam literature:
\begin{enumerate}[label=(\roman*)]
    \item we complete the numerical test of the asymptotic formula \cite{barrettAsymptoticAnalysisEnglePereiraRovelliLivine2009} for a Lorentzian 4-simplex initiated in \cite{donaNumericalStudyLorentzian2019};
    \item following \cite{donaNumericalAnalysisSpin2020}, we compute the partition function of the $\dtre$ graph --- a spin foam with 3 vertices and one internal face --- using the BF and the EPRL vertices, and we find evidence of the emergence of the so-called ``flatness problem'' in both cases.
\end{enumerate}
Along with the discussion and the results, we provide various snippets of the code used in this work to show some concrete usage examples. Finally, in Section \ref{sec:conclusions} we summarize our results and we propose many interesting applications for our code. The computations in the present work have been performed on a laptop with 4 physical cores and on the Centre de Calcul Intensif d'Aix-Marseille using up to 1500 cores. \codek{sl2cfoam-next} is open source and can be accessed online from a public repository \cite{gozziniSL2CfoamnextComputingEPRL}, along with documentation for compiling and installing the library.


\section{The Lorentzian EPRL model}
\label{sec:eprl}

Spin foams provide a regularization of the gravitational path integral over simplicial complexes. The regularized partition function over a 2-complex $\mathcal{K}^*$ made by vertices, edges and faces can be written as
\begin{equation}
\label{eq:partf}
Z_{\mathcal{K}^*} = \sum_{j_f, i_e} \prod_f A_f(j_f) \prod_e A_e(i_e) \prod_v A_v(j_f, i_e) 
\end{equation}
with spin labels coloring faces $f$ and intertwiner labels coloring edges $e$. The functions $A_f, A_e$ and $A_v$ are, respectively, the face amplitude, edge amplitude and vertex amplitude of the chosen model. It is possible to associate a geometrical interpretation to the sum by considering the simplicial complex $\mathcal{K}$ dual to $\mathcal{K}^*$: a vertex in $\mathcal{K}^*$ is dual to a 4-simplex in $\mathcal{K}$, an edge is dual to a tetrahedron and a face is dual to a triangle. 

The EPRL model is a spin foam model which weakly implements at the quantum level the linear simplicity constraint of the Plebanski action for general relativity \cite{engleLQGVertexFinite2008,freidelNewSpinFoam2008}. For our purposes, it is best to consider the formulation of \cite{spezialeBoostingWignerNjsymbols2017}, and absorb the edge amplitude function into the vertex amplitude. The partition function we consider here is
\begin{equation}
\label{eq:partf-2}
Z_{\mathcal{K}^*} = \sum_{j_f, i_e} \prod_f (2j_f + 1) \prod_v A_v(j_f, i_e)
\end{equation}
where the vertex amplitude $A_v(j_f, i_e)$ can be expanded as the sum over additional ``virtual'' spins and intertwiner labels as
\begin{equation}
\label{eq:vertex-ampl}
A_v(j_f, i_e) = \left(\, \prod_{e=1}^5 \,\sqrt{2i_e + 1} \right) \sum_{l_f, k_e} \left(\, \prod_{e=2}^5 \,(2k_e + 1)\, B_4^\gamma(j_f, i_e; l_f, k_e) \;\right) \{15j\}(l_f, k_e)
\end{equation}
where the edge with label $e=1$ is excluded from the product because of redundancy\footnote{We note here the following two differences with the old version \codek{sl2cfoam}: the gauge-fixed intertwiner index and the overall normalization factor of the vertex amplitude.}. We use the labeling $j_{ab}$ or $l_{ab}$ with $a < b$ for the strand that connects nodes $a$ and $b$ on the boundary of the vertex (the nodes are where the vertex edges intersect a fictitious 3-sphere that encloses the vertex). The labels $i_e$ or $k_e$ denote the intertwiners along the edge $e$, with $i_e$ on the boundary. According to \eqref{eq:vertex-ampl}, the virtual spins $l_{12}, l_{13}, l_{14}$ and $l_{15}$ are fixed to the same values of $j_{12}, j_{13}, j_{14}$ and $j_{15}$ that comes from the  boundary of the vertex by the gauge-fixing. The virtual spins $l_f \geq j_f$ have infinite range, and \eqref{eq:vertex-ampl} is a convergent sum \cite{engleRegularizationFinitenessLorentzian2009} in the $l_f$ indices. The case $l_f = j_f$ has been called the \emph{simplified} EPRL model. The parameter $\Delta s$ that sets the homogeneous cutoff $j_f \leq l_f \leq j_f + \Delta s$ on the virtual spins is denoted as the ``number of shells'' in the following. 

Following \cite{spezialeBoostingWignerNjsymbols2017,donaNumericalMethodsEPRL2018}, we call the symbols $B_4^\gamma$ \emph{booster functions} and we define them as
\begin{align}
\label{eq:booster}
B_4^\gamma(j_1, \ldots, j_4, i; l_1, \ldots, l_4, k) &= \sum_{m_i} \Wfour{j_1}{j_2}{j_3}{j_4}{m_1}{m_2}{m_3}{m_4}{i} \Wfour{l_1}{l_2}{l_3}{l_4}{m_1}{m_2}{m_3}{m_4}{k} \\
&\times \int_0^\infty \dd r\ \frac{\sinh^2 r}{4\pi}\ d^{\gamma(j_1+1), j_1}_{j_1l_1m_1}(r)\ d^{\gamma(j_2+1), j_2}_{j_2l_2m_2}(r)\ d^{\gamma(j_3+1), j_3}_{j_3l_3m_3}(r)\ d^{\gamma(j_4+1), j_4}_{j_4l_4m_4}(r) \nn
\end{align}
where we introduced two Wigner $4jm$-symbols on the first line and the boost matrix elements $d^{\rho,k}_{jlm}$ of $\SL$ in the integrand of the second line. Notice that, slightly unconventionally, we define the so-called Y-map on $\SL$ irreps as $(\rho, k) \mapsto (\gamma(j+1), j)$. It can be shown that this choice implements the linear simplicity constraints exactly and not only in the large $j$ limit \cite{alexandrovNewVerticesCanonical2010,dingVolumeOperatorCovariant2010}.

The symbol $\{15j\}(j_f, i_e; l_f, k_e)$ is the $15j$-symbol of first type from the recoupling theory of $\SU$ \cite{yutsisMathematicalApparatusTheory1962}. It can be written as
\begin{align}
\label{eq:15j}
    \{15j\}(j_f, i_e; l_f, k_e) &= \left \{ \begin{array}{ccccc} i_1 & j_{14} & k_4 & l_{24} & k_2 \\[3pt] j_{15} & l_{45} & l_{34} & l_{23} & j_{12} \\[3pt] l_{25} & k_5 & l_{35} & k_3 & j_{13} \end{array}\right \} \\[5pt]
     &= \sum_x \,(2x+1)\, (-1)^{i_1 + k_2 + \cdots + j_{12} + \cdots + l_{23} + \cdots}\, \nonumber \\[4pt] 
   &\times \, \Wsix{i_1}{l_{25}}{x}{k_5}{j_{14}}{j_{15}} \, \Wsix{j_{14}}{k_5}{x}{l_{35}}{k_4}{l_{45}} \Wsix{k_4}{l_{35}}{x}{k_3}{l_{24}}{l_{34}} \nn \\[4pt] 
   &\times \,\Wsix{l_{24}}{k_3}{x}{j_{13}}{k_2}{l_{23}} \Wsix{k_2}{j_{13}}{x}{i_1}{l_{25}}{j_{12}} \nn
\end{align}
where the symbols in the last two lines are the usual $6j$-symbols of recoupling theory of three angular momenta. 

We refer to \cite{spezialeBoostingWignerNjsymbols2017, donaNumericalMethodsEPRL2018,donaNumericalStudyLorentzian2019} for more details on the splitting and the approximation in the number of shells. In particular, we refer to \cite{donaNumericalMethodsEPRL2018} for a nice graphical representation of the vertex amplitude and the sum over virtual spins and intertwiners. Using the graphical representation, the composition of various vertex amplitudes to produce e.g.\ the partition function \eqref{eq:partf-2} can be visualized as the juxtaposition of the various symbols respecting the connections of the corresponding simplicial complex.




\section{The \lib\ library}
\label{sec:library}

The first release of the library \texttt{sl2cfoam} \cite{donaNumericalMethodsEPRL2018} has proved extremely valuable in starting the field of numerical spin foam simulations. Nevertheless, it has many technical limitations that prevented extensions of the simple models considered so far. In this section we describe how using techniques from the field of High Performance Computing it is possible to realize a major step forward in performance and extend considerably the ensemble of models that can be simulated. The new and completely rewritten version of the library is called \lib{}. The main code is written in C and there are Julia bindings for interactive use. The code depends on a number of external libraries for the arbitrary precision routines, parallelization, matrix algebra and other tasks. For the efficient computation of the Wigner symbols we use the \codek{WIGXJPF} and \codek{FASTWIGXJ} libraries \cite{johanssonFastAccurateEvaluation2015}.

\subsection{Internals}

\subsubsection*{Booster coefficients}

The booster coefficients \eqref{eq:booster} are computed by numerically integrating the product of the $\SL$ boost matrix elements and then performing the contraction with the $4jm$-symbols over the $m_i$ indices. By the change of variable $r \mapsto e^{-r}$ the integral is mapped to the finite range $(0\; 1]$. The integrand is an highly oscillatory function depending on the values of $j_f$ and $l_f$ (this comes from the exponential squeezing of larger and larger intervals of the real line towards 0 by the map $r \mapsto e^{-r}$). For an accurate numerical evaluation of this singular integrands, the interval $[0\;1]$ is divided in $n$ subintervals, where $n$ is chosen large enough according to the values of $j_f, l_f$, the Barbero-Immirzi constant $\gamma$ and a global \codek{accuracy} parameter . Defining $I_1 = [0\; \Delta x]$, the next intervals are stretched proportionally as $|I_2| = 2|I_1|, |I_3| = 3|I_1|, \ldots, |I_n| = n|I_1|$. Solving for $\Delta x$ then $\Delta x = 2 / (n(n+1))$ so that the size of the subintervals decreases while approaching 0, where more precision is required for the numerical integration routine. Each subinterval is integrated numerically using Gauss-Kronrod quadrature with 30 and 61 points in double or quadruple (128 bits) precision. The result is the sum over all subintervals of the quadrature with the Kronrod points and an estimate of the error is provided by the difference with the sum over quadratures computed with the Gauss points. This provides an additional check that the error in the numerical integration is kept small.

The functions $d^{\rho k}_{jlm}(r)$ are computed in an efficient way as finite sums of complex exponentials, using the following formula \cite{colletSimpleExpressionUnitaryirreducible2018}
\begin{align}
\label{eq:francois}
    d^{\rho k}_{jlm}(r) &= \frac{1}{(e^r - e^{-r})^{j+l+m}} \; \times \\[5pt]
 &\quad\left[ \sum_{a=0}^{j+l - |k-m| } Y_a^{(\rho k)jlm} e^{r(j+l- |k-m| -2a -\I \rho)} + (-1)^{l-j} \sum_{b=0}^{j+l - |k+m| } \overline{Y_b^{(\rho k)lj-m}} e^{r(j+l- |k+m| -2b +\I \rho)} \right] \nn
\end{align}
where $Y_a^{(\rho k)jlm}$ are complex coefficients and $\rho \neq 0$. The formula suffers from catastrophic cancellation of most of the significant digits already at spins as low as $\sim 5$, using standard double-precision floating-point variables. For this reason, it is necessary to compute all the terms using arbitrary precision arithmetic, with a number of bits that increases as the average spin labels increase. The code that computes \eqref{eq:francois} has been rewritten from scratch to increase the performance and the stability with respect to previous versions. In particular, the new code does not suffer from the instabilities reported in \cite{donaNumericalStudyLorentzian2019} for spins above $\sim 50$.

If the shell parameter $\Delta s$ is greater than zero, it is necessary to compute a large number of booster coefficients and store them. Since one of the four $l_f$ indices on each edge is always gauge-fixed to a vertex boundary spin $j_f$, the number of booster coefficients to compute is $(\Delta s + 1)^3$ times the number of possible intertwiners $(i,k)$ that varies depending on the $l_f$ indices. It is convenient to combine all these coefficients into multidimensional arrays as explained in the next section.

\subsubsection*{Tensor contractions}

Computing a general partition function as \eqref{eq:partf} requires to perform a huge number of sums and products. The most efficient way\footnote{Parallelization is treated in the next subsection. Here we consider a single processing unit.} of performing these operations on computer hardware is to use specialized routines (such as the BLAS standard) that are optimized for computing matrix products. These can be used also for computing more general expressions, such as the contraction over one index of multidimensional arrays with many indices --- we call them \emph{tensors}\footnote{The term \emph{tensor} is used here as in computer science, where it refers to multidimensional arrays.}. For example, consider the following contraction of two tensors:
\[
Z^{abde} = \sum_k X^{abk} Y^{kde}
\]
over index $k$. Regrouping the outer indices as $(ab) = I, (de) = J$, the expression becomes
\[
\sum_k X^{Ik} Y^{kJ} = (X \cdot Y)^{IJ} = Z^{(ab)(de)}
\]
where the middle product is a matrix product that can be computed using specialized routines. The procedure can be iterated for contracting over all the required indices\footnote{This is sometimes called ``Loop-over-GEMM'' from the standard term GEMM for BLAS matrix product. There exist more specialized schemes for tensor contractions but they are either less flexible or very complex to implement.}. Note that it is important to respect the layout of data in memory and that in general, among all the possible ways of contracting many indices, there are certain combinations of steps that are more efficient than others --- mainly, one wants to reduce to the minimum the operations of transposition of the data that are required to align the indices in the correct order for the optimized matrix product. Note also that in practice one is limited to consider a ``reasonable'' number of indices, since a tensor with $n$ indices, each one with an average of $k$ possible values, has exponential dimension of $k^n$, and the contraction of two such tensor is multiplicative in the dimension:
\[
[k^n] \cdot [k^n] \to [k^{(n-1)(n-1)}] \approx [k^{n^2}].
\]
Since a double-precision number requires 8 bytes of memory to be stored in hardware, a ``small tensor'' with e.g. 6 indices in $(1,\ldots,10)$ requires around 8MBs of memory, but the contraction of two such tensors over one index requires a considerably larger amount of 80GBs of memory.

The spin foam partition function \eqref{eq:partf-2} contains the vertex amplitude $A_v$ as the main building block. The sum is over spin labels on the bulk faces and intertwiner labels on the bulk edges.  A single vertex boundary is completely defined by the spin labels of the 10 boundary faces and of the 5 boundary intertwiners. If the spins on the faces are held fixed, the partition function is the contraction of all vertex amplitudes over all the internal intertwiner indices. Therefore, it is convenient to represent a vertex in hardware as a 5-dimensional tensor in the 5 intertwiner indices $(i_1k_2k_3k_4k_5)$, parametrized by the fixed parameters $(j_{12}j_{13} \cdots)$ of the boundary faces. The complete partition function then is computed as the outer sum over all bulk faces of the contraction over all bulk edges. Writing the vertex tensor as $A^{i_1i_2i_3i_4i_5}_{j_a}$, the sum becomes
\begin{equation}
\label{eq:vertex-contraction}
Z_{\mathcal{K}^*} = \sum_{j_f} A^{i_1i_2i_3i_4i_5}_{j_a} A^{i'_1i'_2i'_3i'_4i'_5}_{j_b} \cdots
\end{equation}
where repeated upper indices are contracted according to the connectivity of $\mathcal{K}^*$. From now on we use a similar notation for all tensor symbols, with ordered upper indices labeling all the running tensor indices and lower indices denoting external parameters that are fixed for the tensor.

A similar strategy can be used to compute the fundamental vertex tensor $A$. By \eqref{eq:vertex-ampl} it is the contraction of the booster functions $B_4^\gamma$ and the $15j$-symbols over the 6 virtual spins $(l_{23}l_{24}l_{25}l_{34}l_{35}l_{45})$ and the 4 virtual intertwiners $(k_2k_3k_4k_5)$. Actually, it is most efficient to build tensors using also the $(l_f)$ indices. 
A single booster tensor can be written as
\begin{equation}
B_{j_1j_2j_3j_4}^{l_1l_2l_3l_4ik}
\end{equation}
and a $15j$-symbol tensor as
\begin{equation}
\label{eq:15j-contraction}
\{15j\}_{j_{12}j_{13}j_{14}j_{15}}^{l_{23}l_{24}l_{25}l_{34}l_{35}l_{45}i_1k_2k_3k_4k_5}.
\end{equation}
By the above remarks, it is not practical to compute a full tensor with 11 indices if the average index is of order $10$ or more. Therefore, the computation of the vertex tensor is split in the following steps:

\medskip
\begin{framed}
\begin{enumerate}
\item Input: 10 spins $(j_{12}j_{13} \cdots)$ and number of shell $\Delta s$
\item Compute the ranges of the intertwiner indices $(i_1i_2i_3i_4i_5)$
\item Boosters: compute the following 4 tensors
\[
B_{j_{23}j_{24}j_{25}j_{12}}^{l_{23}l_{24}l_{25}j_{12}i_2k_2}, 
B_{j_{34}j_{35}j_{13}j_{23}}^{l_{34}l_{35}j_{13}l_{23}i_3k_3}, 
B_{j_{45}j_{14}j_{24}j_{34}}^{l_{45}j_{14}l_{24}l_{34}i_4k_4}, 
B_{j_{15}j_{25}j_{35}j_{45}}^{j_{15}l_{25}l_{35}l_{45}i_5k_5}
\]
\item $15j$-symbol: compute the following 5 tensors for $6j$-symbols as in \eqref{eq:15j}
\[
S_{j_{14}j_{15}}^{l_{25}i_1k_5x}, S_{j_{14}}^{l_{35}l_{45}k_4k_5x}, S^{l_{24}l_{34}l_{35}k_3k_4x}, S_{j_{13}}^{l_{23}l_{24}k_2k_3x}, S_{j_{12}j_{13}}^{l_{25}i_1k_2x}    
\]
\item Assembly: loop over $(l_{23}l_{24}l_{25}l_{34}l_{35}l_{45})$ and
\item %
\begin{enumerate}[label=(\roman*)]
    \item Loop over $(i_1k_2k_3k_4k_5)$ and $x$ and compute \eqref{eq:15j-contraction} as a 5-dimensional tensor $W^{i_1k_2k_3k_4k_5}$
    \item Select 4 submatrices $B^{i_ak_a}$ in the $B$ tensors according to current $(l_{23}l_{24}l_{25}l_{34}l_{35}l_{45})$
    \item Contract the booster submatrices with $W$ over virtual intertwiner indices $(k_2k_3k_4k_5)$ 
    \item Accumulate 
\end{enumerate}
\item Result: the vertex tensor $A^{i_5i_4i_3i_2i_1}_{j_a}$
\end{enumerate}
\end{framed}
\medskip

Note that in the result the vertex tensor indices are reversed\footnote{This comes from the fact that it is convenient to split large tensors over the gauge-fixed index, which is $i_1$, and due to the chosen memory layout (column-major, for binary bitwise interoperability between the C library and the Julia interface), the juxtaposition in memory of binary arrays corresponds to extension of the rightmost index.}. The triangular inequalities are verified at each step to restrict the number of loops. In case the final vertex tensor happens to be too large to fit in working memory, the computation is split in batches of smaller tensors that are joined at the end. It is also possible to compute the vertex tensor for a reduced set of intertwiner indices.

\subsubsection*{Parallelization}

The library is extensively parallelized using an hybrid OpenMP-MPI scheme. In case the shell parameter $\Delta s$ is not zero, the parallelization is performed over shells, i.e.\ over the virtual spins $l_f$. This is done both for the computation of the booster tensors and for the vertex tensor. For the former, the total number of loops over $l_f$ is $N_B = (\Delta s + 1)^3$, while for the latter it is $N_A = (\Delta s + 1)^6$. When the computation is launched on multiple MPI nodes, the $N_B$ and $N_A$ loops are distributed among all the nodes. Each node then parallelizes its own loops across its CPUs using OpenMP. If the number of assigned loops to one node is too low, so that this strategy will cause some CPUs to remain idle, the parallelization in each node is moved from shells to inner loops: for the boosters, the sums of \eqref{eq:booster} are parallelized, while for the vertex the parallelization is over intertwiners $(i_1k_2k_3k_4k_5)$ in assembling the internal $15j$-tensor.

The contraction of vertex tensors to compute an amplitude with many vertices can be parallelized as well. If there are bulk faces to be summed over, it is convenient to parallelize over these sums. It is also possible to use automatic parallelization while performing tensor contractions using BLAS libraries such as MKL or OpenBLAS. Additionally, tensor contractions can be offloaded to the GPU and parallelized over thousands of GPU cores using the Julia module \codek{SL2CfoamGPU} presented in the next section. 


\subsection{Interface and examples}

The library provides a native C interface and also Julia bindings for interactive use. Both interfaces provide methods to:
\begin{itemize}
    \item initialize and configure the library
    \item compute and load EPRL vertex tensors
    \item compute and load booster tensors
    \item compute BF vertex ($15j$-symbol) tensors
    \item compute $B_4^\gamma$ coefficients
    \item compute Livine-Speziale coherent state coefficients
\end{itemize}

All the tensors and coefficients can be contracted according to the connectivity of the spin foam to obtain the total amplitude. The Julia interface provides various methods \codek{contract} to perform the contractions. These are fast operations compared to computing the amplitude tensors, but they might nevertheless have a considerable impact on the computational time if a large number of contractions is required, for example in a spin foam with many vertices or in contracting tensors with a large number of entries. To speed up this part of the computation, it is possible to offload the tensor contractions to the GPU. This is handled transparently in the \codek{contract} functions using \codek{CUDA.jl} \cite{besardEffectiveExtensibleProgramming2019}, which in turn relies on optimized routines from the CuBLAS library \cite{CuBLAS}. We show an example of GPU offloading in the next section.

There are a number of options for controlling the library that can be set at library initialization. Importantly, there are three levels of accuracy for the computation of the booster coefficients: \codek{NormalAccuracy}, \codek{HighAccuracy} and \codek{VeryHighAccuracy}. As a rule of thumb, values of the Immirzi parameter $\gamma \gg 1$ or computations involving a huge number of vertex tensors require higher accuracy. The Immirzi parameter must be set at initialization but can be changed later. In Listing \ref{code:init} we show how to initialize the library. The options to be passed include the Immirzi parameter, a working folder with precomputed tables of $3j$ and $6j$-symbols, and other parameters for controlling the memory usage.
\begin{jllisting}[caption={Initialization of the library.}, label={code:init}]
    using SL2Cfoam
    using HalfIntegers

    # init SL2Cfoam library
    Immirzi = 0.123;
    folder = "/path/to/working/folder";
    conf = SL2Cfoam.Config(VerbosityOff, NormalAccuracy, 100, 0);
    SL2Cfoam.cinit(folder, Immirzi, conf);
\end{jllisting}
In Listing \ref{code:vertex-compute} we show how to compute a vertex tensor. The command \codek{vertex\_compute} takes as input the list of the 10 boundary spins of the vertex and the number of shells, and outputs the vertex tensor. Optionally, it is possible to compute only a restricted range of boundary intertwiners.
\begin{jllisting}[caption={Computation of a vertex tensor.}, label={code:vertex-compute}]
    spins = [2 2 2 2 2 2 2 2 2 2]
    shells = 1
    v = vertex_compute(spins, shells);
    
    # show the amplitude with intertwiners (0,0,0,0,0)
    # v.a is the array with the data
    @show v.a[1,1,1,1,1];
\end{jllisting}
In Listing \ref{code:contract} we show how to define 5 coherent states with random angles and then contract the vertex with them to produce a coherent amplitude.
\begin{jllisting}[caption={Computation of a coherent amplitude with random angles.}, label={code:contract}]
    # define the 4 spins that enter the coherent state
    js_intw = [2 2 2 2]
    
    # compute the coherent states with random angles in (0, pi)
    # angles are 4x2 matrices (theta, phi) [ 2 angles per 4 normals ]
    cs1 = coherentstate_compute(js_intw, pi * rand(4,2));
    cs2 = coherentstate_compute(js_intw, pi * rand(4,2));
    cs3 = coherentstate_compute(js_intw, pi * rand(4,2));
    cs4 = coherentstate_compute(js_intw, pi * rand(4,2));
    cs5 = coherentstate_compute(js_intw, pi * rand(4,2));
    
    # contract the vertex with the coherent states (notice the order)
    contract(v, cs5, cs4, cs3, cs2, cs1)
\end{jllisting}
Finally, in Listing \ref{code:simple-asym} we show how to perform a simple asymptotic analysis of a Euclidean vertex with all spins $j_1 = \lambda, \lambda = 1 \to 10$ and $\Delta s=1$. In this simple example, the amplitude is exponentially suppressed in $\lambda$ since the random angles do not satisfy the closure constraint.
\begin{jllisting}[caption={Simple asymptotics.}, label={code:simple-asym}]
# compute vertices
# all boundary spins set to λ (Euclidean boundary configuration)
@time vs = [ vertex_compute(λ * ones(10), 1) for λ in 1:10 ]

# compute and contract with random coherent states
ampls = ComplexF64[]
for λ = 1:10
    
    css = [ coherentstate_compute(λ * ones(HalfInt, 4), pi * rand(4,2)) for i in 1:5 ]
    ampl = contract(vs[λ], css...)
    push!(ampls, ampl)
    
end

# logplot of absolute value of the coherent amplitudes
using Plots
plot(1:10, abs.(ampls), yaxis=:log, shape=:circle, title="Simple asymptotics (suppressed).")
\end{jllisting}
The combined code from the previous examples can be easily run on a consumer laptop in a few seconds. We refer to the online repository and to the source code \cite{gozziniHighPerformanceCode2021} for extensive comments on all the available methods and options. 

\subsection{Benchmarks}


We report the results of some simple benchmark to provide an estimate of the improvements with respect to previous version \codek{sl2cfoam} and of the scaling the computational time with varying parameters. We provide also some snippets of the code used for the benchmarks in order to show more examples on how to use the Julia interface.

\subsubsection*{Booster coefficients}

\begin{figure}[h]
    \centering
    \begin{subfigure}[b]{0.48\textwidth}
        \centering
        \includegraphics[width=\textwidth]{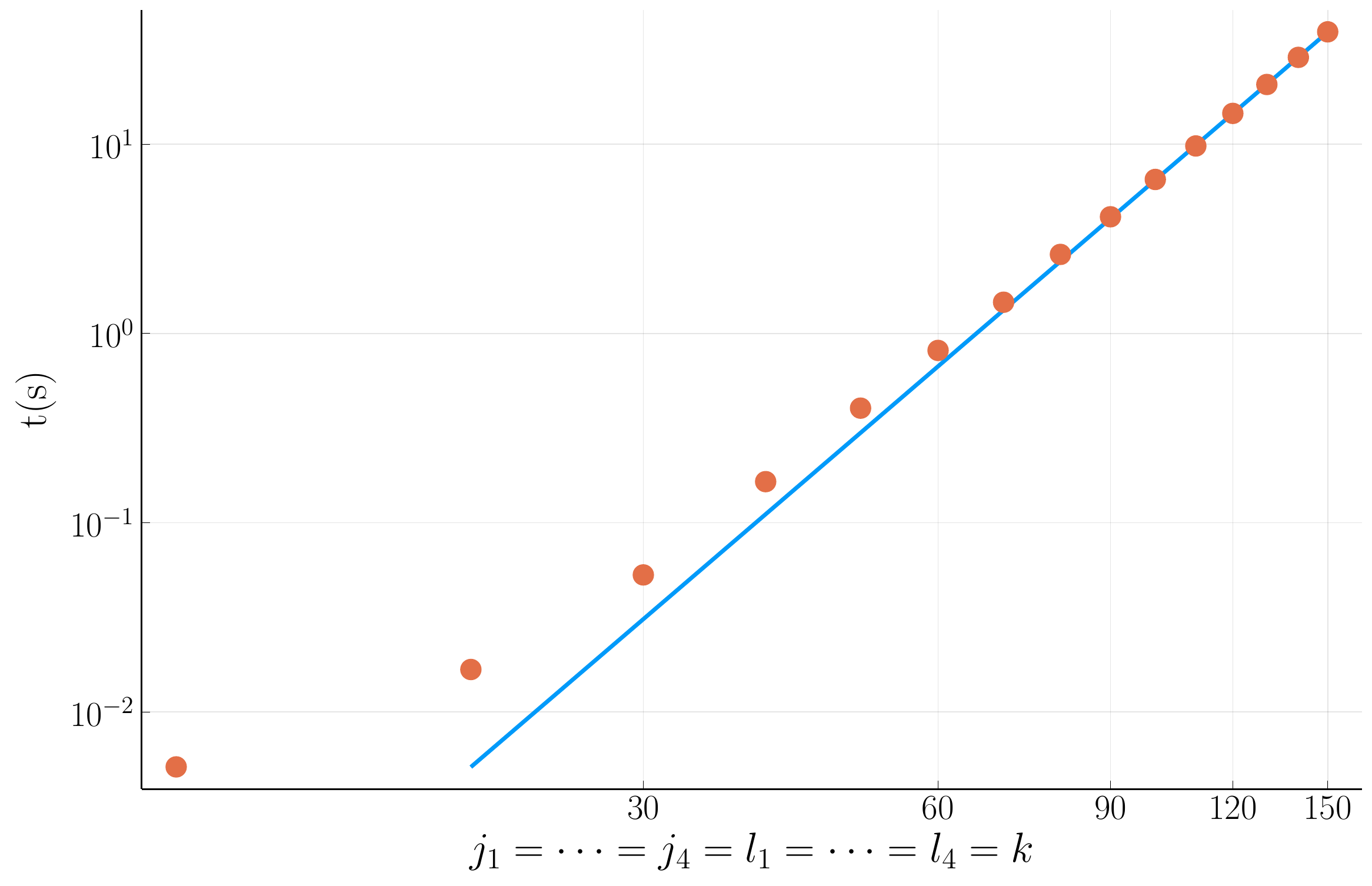}
    \end{subfigure}
    \hfill
    \begin{subfigure}[b]{0.48\textwidth}
        \centering
        \includegraphics[width=\textwidth]{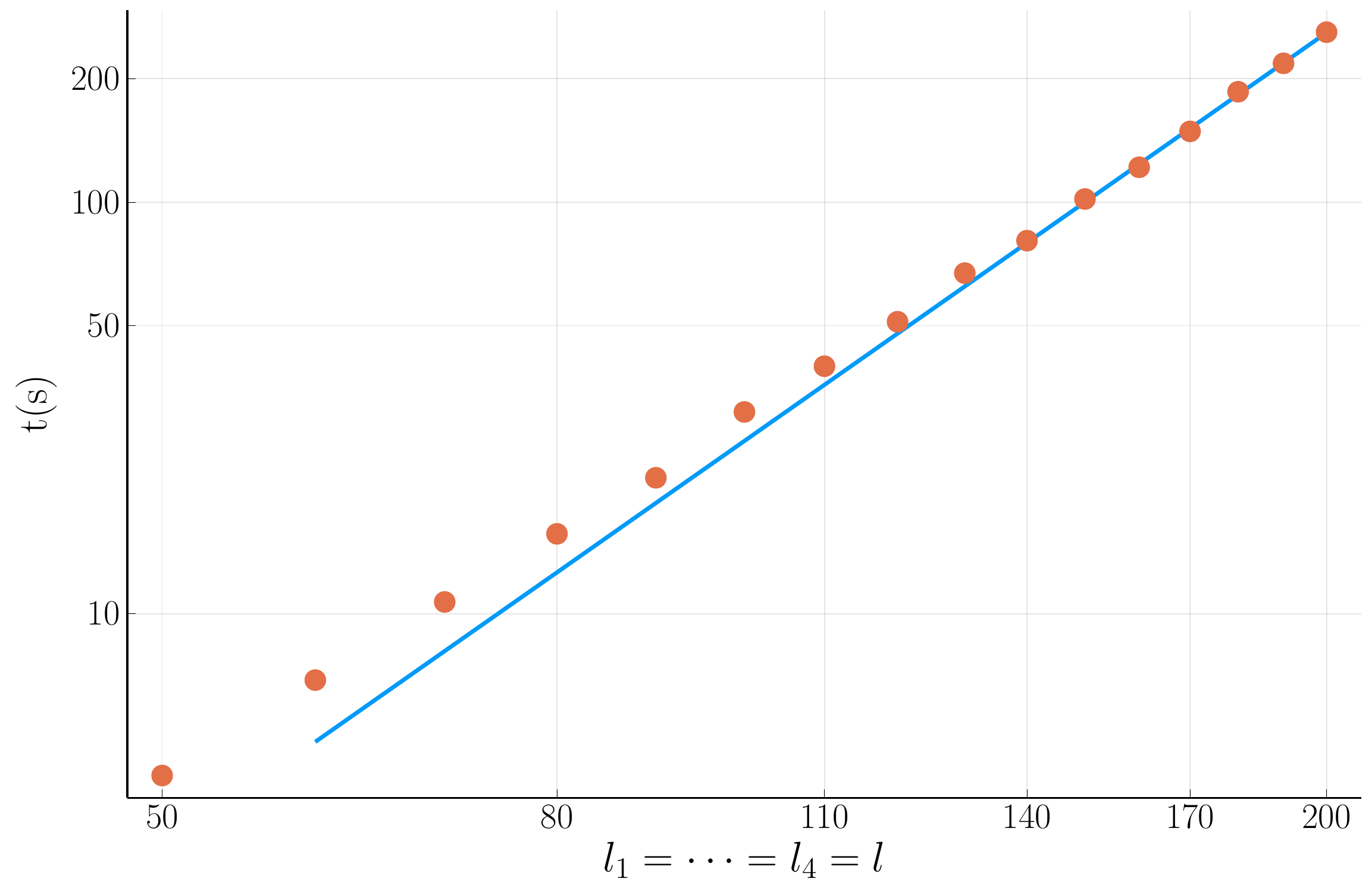}
    \end{subfigure}
    \caption{Log-log plots of times in seconds for computing booster coefficients. The computation run on a server with 32 cores. \emph{Left:} Minimal case with increasing uniform spin $j_i = l_i = k$. The fitting power law has exponent $4.4$. \emph{Right:} Non-minimal case with $j_1 = \cdots = j_4 = 50$ and increasing spins $l_1 = \cdots = l_4 = l$. The fitting power law has exponent $3.3$.}
    \label{fig:booster-times}
\end{figure}

The computation of booster coefficients was one of the major slowdowns of the old version of the code. Moreover, the previous code suffered from instabilities kicking in at spins of the order $\sim 50$ which prevented the authors to go beyond the value $\lambda=9$ in the Lorentzian asymptotic analysis \cite{donaNumericalStudyLorentzian2019}. The rewritten code for the booster coefficients does not suffer from instabilities and is more than 2 orders of magnitude faster on a single core. We report in Table \ref{tab:booster-new-vs-old} some simple time comparisons on a laptop with 4 cores. A study of the scaling of computational time with varying spin on a server with 32 cores is presented in Figure \ref{fig:booster-times}. The code for the first of the two tests of Figure \ref{fig:booster-times} is reported in Listing \ref{code:bench-boost}.
\begin{jllisting}[caption={Boosters benchmark.}, label={code:bench-boost}]
for K = 1:15

    t = @elapsed b4_compute(K .* [10 10 10 10], K .* [10 10 10 10])
    println("time for K = $K : $t")

end
\end{jllisting}
{\renewcommand{\arraystretch}{1.5}
\begin{table}[h]
    \centering
    \begin{tabular}{ ||c|c|c|c|c|| } 
     \hline
     $j = 5$ & $j = 10$ & $j = 15$ & $j = 20$ & $j = 25$\\ 
     \hline
     50x & 70x & 100x & 130x & 180x \\ 
     \hline
    \end{tabular}
    \caption{Approximate increase in performance (computed as the ratio \emph{time[\codek{sl2cfoam}]} over \emph{time[\codek{sl2cfoam-next}]}) for computing a minimal booster $B_4^\gamma(j,j,j,j,\cdot;j,j,j,j,\cdot)$. Lowest accuracy has been set in both libraries. The computation run on a laptop with 4 cores.}
    \label{tab:booster-new-vs-old}
\end{table}
}
%

\subsubsection*{Vertex tensor}

\begin{figure}[h]
    \centering
    \begin{subfigure}[b]{0.49\textwidth}
        \centering
        \includegraphics[width=\textwidth]{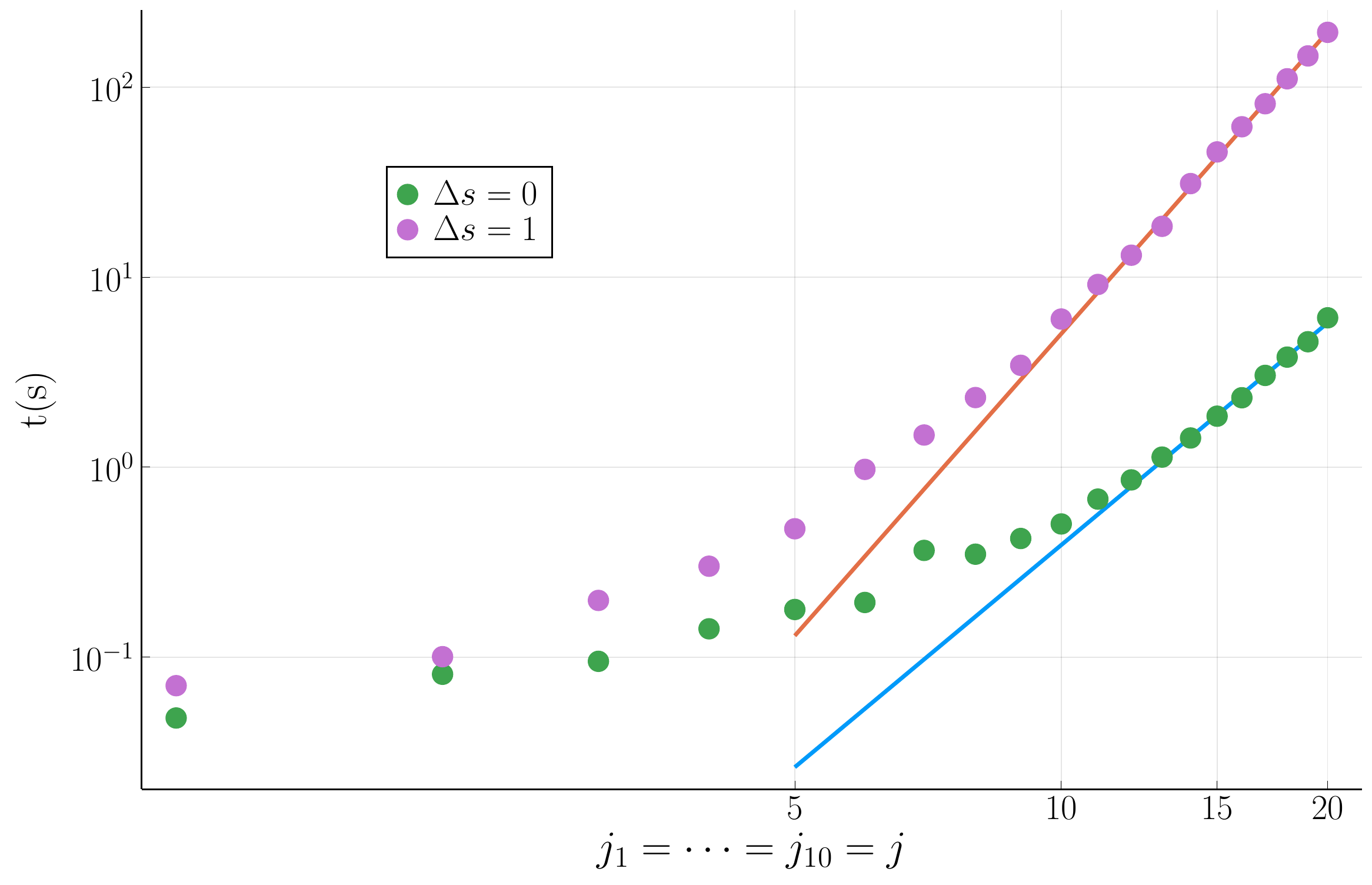}
    \end{subfigure}
    \hfill
    \begin{subfigure}[b]{0.49\textwidth}
        \centering
        \includegraphics[width=\textwidth]{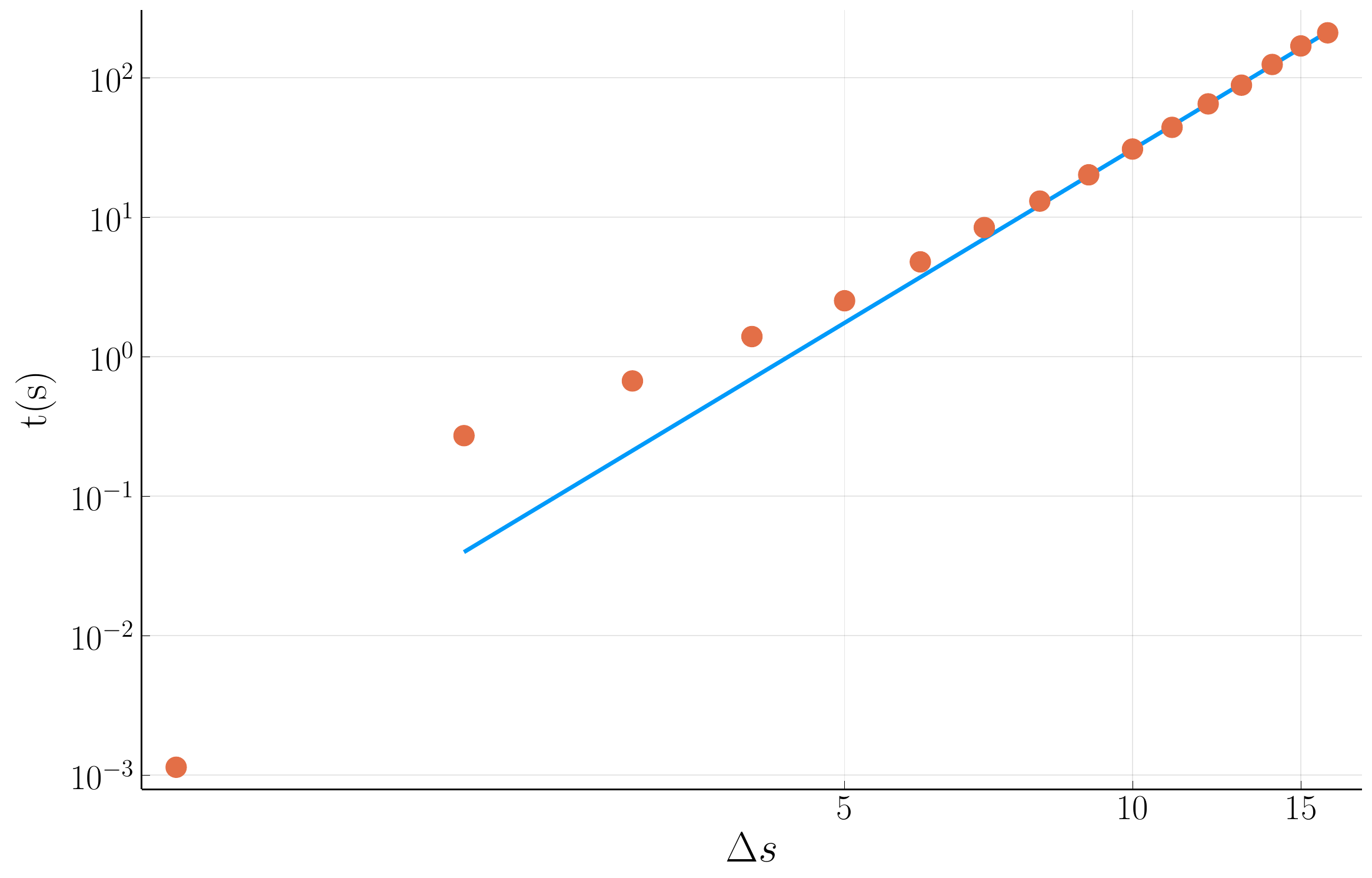}
    \end{subfigure}
       \caption{Log-log plots of times in seconds for vertex tensor computation with increasing boundary spins and number of shells. The computation run on a laptop with 4 cores. \emph{Left:} Increasing boundary spins $j_1 = \cdots = j_{10} = j$ at $\Delta s = 0, 1$. The fitting power laws have exponents $3.9$ for $\Delta s = 0$ and $5.3$ for $\Delta s = 1$. \emph{Right:} Increasing number of shells with fixed boundary spins $j_1 = \cdots = j_{10} = 1$. The fitting power law has exponent $4.1$.}
    \label{fig:ampl-times}
\end{figure}

We compare the time needed to compute a full vertex tensor $A^{i_5i_4i_3i_2i_1}_{j_a}$ with all boundary spins equal $j_a = j$, at number of shells $\Delta s=0$ and $\Delta s = 1$, for increasing values of $j$. We report the results in Figure \ref{fig:ampl-times} and the increase in performance in Table \ref{tab:perf-new-vs-old}. On a laptop the new code is 4 orders of magnitude faster at spins as low a $j=4$ for $\Delta s=1$. Performance further increase going to higher spins and higher number of shells (this without considering using MPI or a GPU, which would provide an additional substantial speedup). The code for the second of the two tests of Figure \ref{fig:ampl-times} is reported in Listing \ref{code:bench-vertex}.
\begin{jllisting}[caption={Vertex benchmark.}, label={code:bench-vertex}]
# this tells the function vertex_compute that for this benchmark
# we don't need neither the resulting tensor nor to store it
benchres = SL2Cfoam.VertexResult((false, false, false))

for shells = 0:16

    t = @elapsed vertex_compute(ones(Int, 10), shells; result = benchres)
    println("time for shells = $shells : $t")

end
\end{jllisting}
{\renewcommand{\arraystretch}{1.5}
\begin{table}[h]
    \centering
    \begin{tabular}{ ||r|c|c|c|c|| } 
     \hline
     \multirow{2}{*}{$\Delta s=0$} & $j=1$ & $j=3$ & $j=5$ & $j=7$ \\ 
     \cline{2-5}
     & 3x & 30x & 1200x & 12000x \\
     \hline\hline
     \multirow{2}{*}{$\Delta s=1$} & $j=1$ & $j=2$ & $j=3$ & $j=4$ \\ 
     \cline{2-5}
     & 50x & 250x & 2000x & 14000x \\ 
     \hline
    \end{tabular}
    \caption{Approximate increase in performance (computed as the ratio \emph{time[\codek{sl2cfoam}]} over \emph{time[\codek{sl2cfoam-next}]}) for computing a full vertex tensor. The computation run on a laptop with 4 cores.}
    \label{tab:perf-new-vs-old}
\end{table}
}

\subsubsection*{Tensor contractions on CPU and GPU}

\begin{figure}[h]
    \centering
    \begin{subfigure}[b]{0.49\textwidth}
        \centering
        \includegraphics[width=\textwidth]{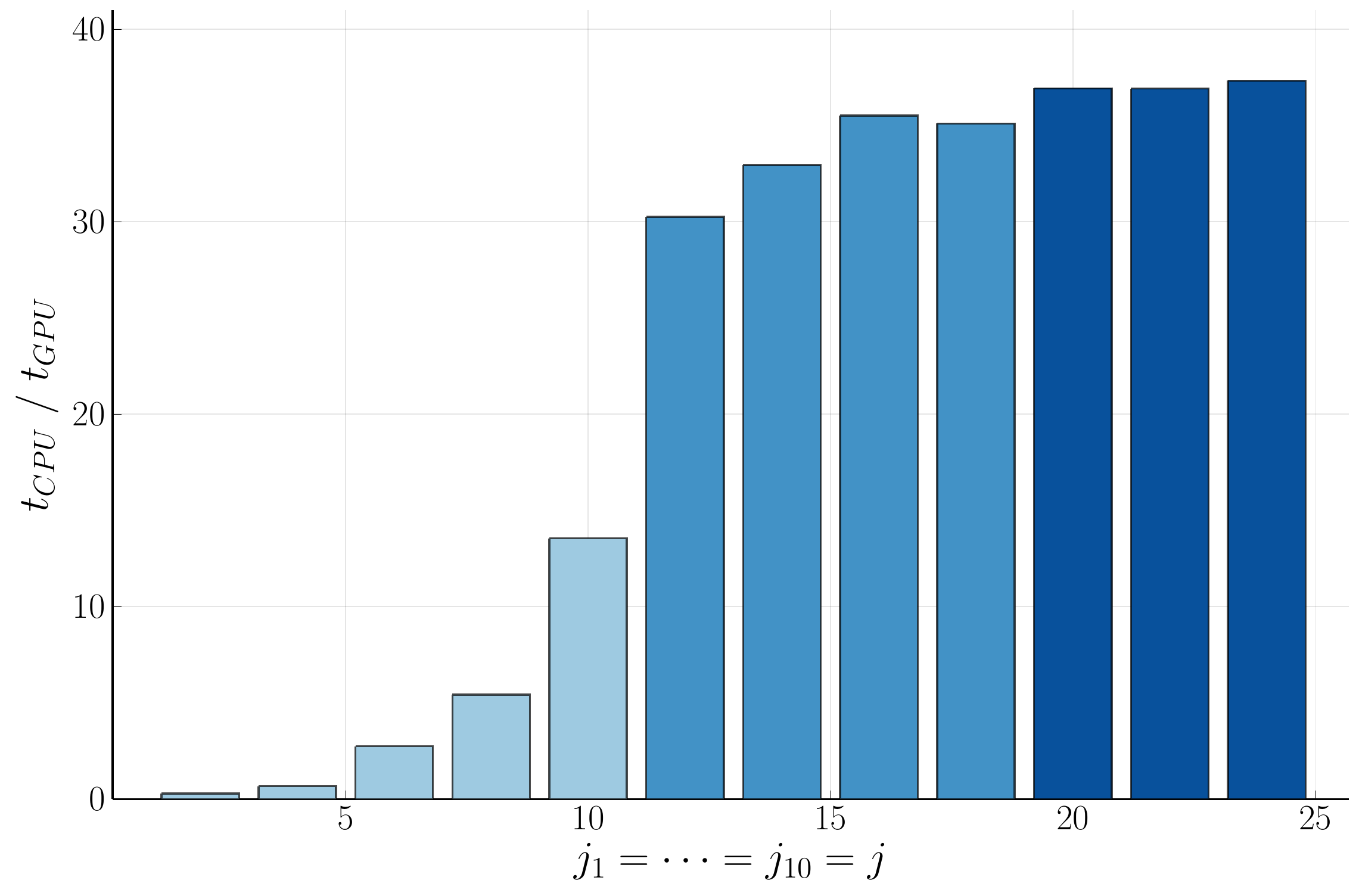}
    \end{subfigure}
    \hfill
    \begin{subfigure}[b]{0.49\textwidth}
        \centering
        \includegraphics[width=\textwidth]{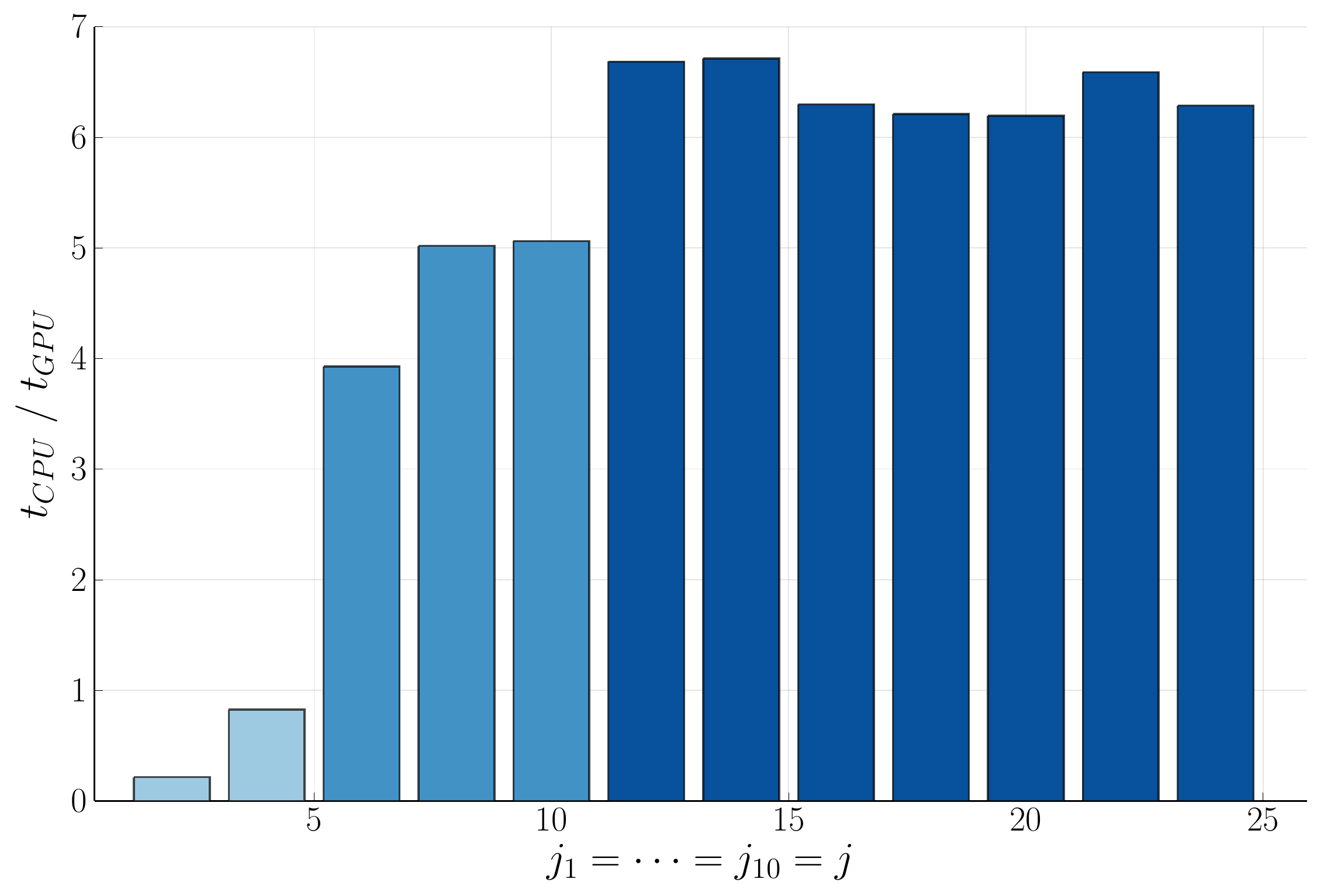}
    \end{subfigure}
       \caption{Performance increase when contracting tensors on the GPU, measured as the ratio between the CPU time and the GPU time. \emph{Left:} Complete contraction of two vertices: $\sum_{i_k} A^{i_5i_4i_3i_2i_1}_{j_a=j}A^{i_5i_4i_3i_2i_1}$. \emph{Right:} Complete contraction of one vertex with 5 coherent states: $\sum_{i_k} A^{i_5i_4i_3i_2i_1}_{j_a=j}\psi^{i_5}_j\psi^{i_4}_j\psi^{i_3}_j\psi^{i_2}_j\psi^{i_1}_j$.}
       \label{fig:contractions}
\end{figure}

Tensor contractions of vertices and coherent states are one of the fastest part of the new code, since they basically boil down to highly optimized matrix multiplications. Indeed, the time spent doing the contractions is almost negligible in scenarios with few vertices as the applications presented in Section \ref{sec:applications}. However, dealing with many vertices it is essential to speedup the contraction process as the number of indices to be summed over increases. In the new code it is possible to offload the contractions to the GPU, using the external library \codek{CUDA.jl} \cite{besardEffectiveExtensibleProgramming2019}, which provides a major speedup. 
As an example, we report in Figure \ref{fig:contractions} the time comparison between CPU and GPU for two different types of contractions. In our simple tests the GPU provided a speedup of contraction time between 6 and 40 times on our hardware configuration (a single Nvidia Tesla P100 card with 12GBs of memory).
The code for the second of the two tests of Figure \ref{fig:contractions} is reported in Listing \ref{code:bench-vertex}. Leveraging the GPU basically requires the single instruction \codek{to\_GPU} to copy the tensor objects to the GPU memory. The contraction is then performed in exactly the same way as for tensors on the CPU.
\begin{jllisting}[caption={GPU benchmark.}, label={code:bench-gpu}]
using SL2CfoamGPU
using CUDA
...
    
for J = 2:2:24

    v = vertex_compute(J * ones(10), 0);
    css  = [ coherentstate_compute([J J J J], pi * rand(4,2)) for i in 1:5 ];
    ct = @elapsed contract(v, css...)
        
    vg = to_GPU(v);
    cssg = [ to_GPU(cs) for cs in css ];
    gt = CUDA.@elapsed contract(vg, cssg...)

    println("ratio of time GPU / CPU for J = $J: $(gt/ct)")
        
end
\end{jllisting}
%


\section{First applications}
\label{sec:applications}

\subsection{Lorentzian 4-simplex asymptotics}


It is known \cite{barrettAsymptoticAnalysisEnglePereiraRovelliLivine2009,barrettLorentzianSpinFoam2010} that the EPRL vertex amplitude oscillates with a phase given by the Regge action in the large spin limit. More precisely, consider a single spin foam vertex with a boundary coherent state whose data correspond to a Lorentzian 4-simplex $\Sigma$. This has been called a \emph{Regge geometry} in the literature. Consider a triangulation of $\Sigma$ made by 5 spacelike tetrahedra $\tau_i$. The Area-Regge action for this triangulation reads 
\begin{equation}
    S_\Sigma = \sum_{a \le b} \theta_{ab} A_{ab}
\end{equation}
where $A_{ab}$ is the area of the triangle shared by tetrahedra $\tau_a, \tau_b$ and $\theta_{ab}$ is the (Lorentzian) 4d angle hinged on this triangle. Let $j_{ab}$ be the spin on the link connecting nodes $a,b$ on the boundary of the spin foam vertex. In the limit for all spins homogeneously large, i.e.\ under a rescaling $j_{ab} \to \lambda j_{ab}$ with $\lambda \to \infty$, the EPRL vertex amplitude has the asymptotic form \cite{barrettAsymptoticAnalysisEnglePereiraRovelliLivine2009,barrettLorentzianSpinFoam2010,donaNumericalStudyLorentzian2019}
\begin{equation}
    \label{eq:lor-asym}
    A_v(j_{ab}, \vec{n}_{ab}) = \frac{(-1)^\chi}{\lambda^{12}} (N_1 e^{i \lambda S_R} + N_2 e^{-i\lambda S_R}) + O(\lambda^{-13})
\end{equation}
up to a global phase factor. In this expression
\begin{equation}
    S_R = \sum_{a \le b} \theta_{ab}(\vec{n}_{ab}) \gamma j_{ab}
\end{equation}
where the Lorentzian angles $\theta_{ab}$ can be computed from the 3d normals $\vec{n}_{ab}$ given as boundary coherent state data. If the areas and normals satisfy the closure and shape-matching constraints \cite{donaSUGraphInvariants2018,donaNumericalStudyLorentzian2019} the area-Regge action $S_R$ is equivalent to the usual Regge action in term of length variables.

The numerical study of the Lorentzian asymptotics \eqref{eq:lor-asym} has been initiated in \cite{donaNumericalStudyLorentzian2019} using the previous version of the code, \codek{sl2cfoam}. The authors computed at various degrees of approximation the amplitude for a vertex with boundary spins
\begin{equation*}
    j_{1a} = 5\lambda \ \ \text{for $a = 2,\ldots,5$} \qquad \text{and} \qquad j_{ab} = 2\lambda \ \ \text{for $1 < a < b$}
\end{equation*}
and boundary normals that reconstruct a Lorentzian 4-simplex with areas corresponding to the given spins $j_{ab}$. Crucially, the performance of the previous code were not sufficient to obtain numerical evidence to support formula \eqref{eq:lor-asym}. The computations could be pushed to the rescaling parameter $\lambda = 9$ with a single shell (i.e.\ $\Delta s = 1)$. This relatively low values did not prove to be enough to see the oscillatory behavior predicted by the exact asymptotics.

Here we report on the results obtained in the same setting using the new version \lib{}. We have been able to push the computation up to values $\lambda \sim 40$ with 6 shells and $\lambda \sim 30$ with 8 shells (using the time and computing resources that have been allocated to this project on our computing facility). The code for generating the following plots is very similar to Listing \ref{code:simple-asym}, with the only difference that the vertex tensors have been computed using MPI with the provided tool \codek{vertex-fulltensor} and then loaded into the Julia code using the function \codek{vertex\_load}. We show an extract in Listing \ref{code:lorasym}.

\begin{figure}[!hbtp]
    \centering
    \begin{subfigure}[b]{1.0\textwidth}
        \centering
        \includegraphics[width=\textwidth]{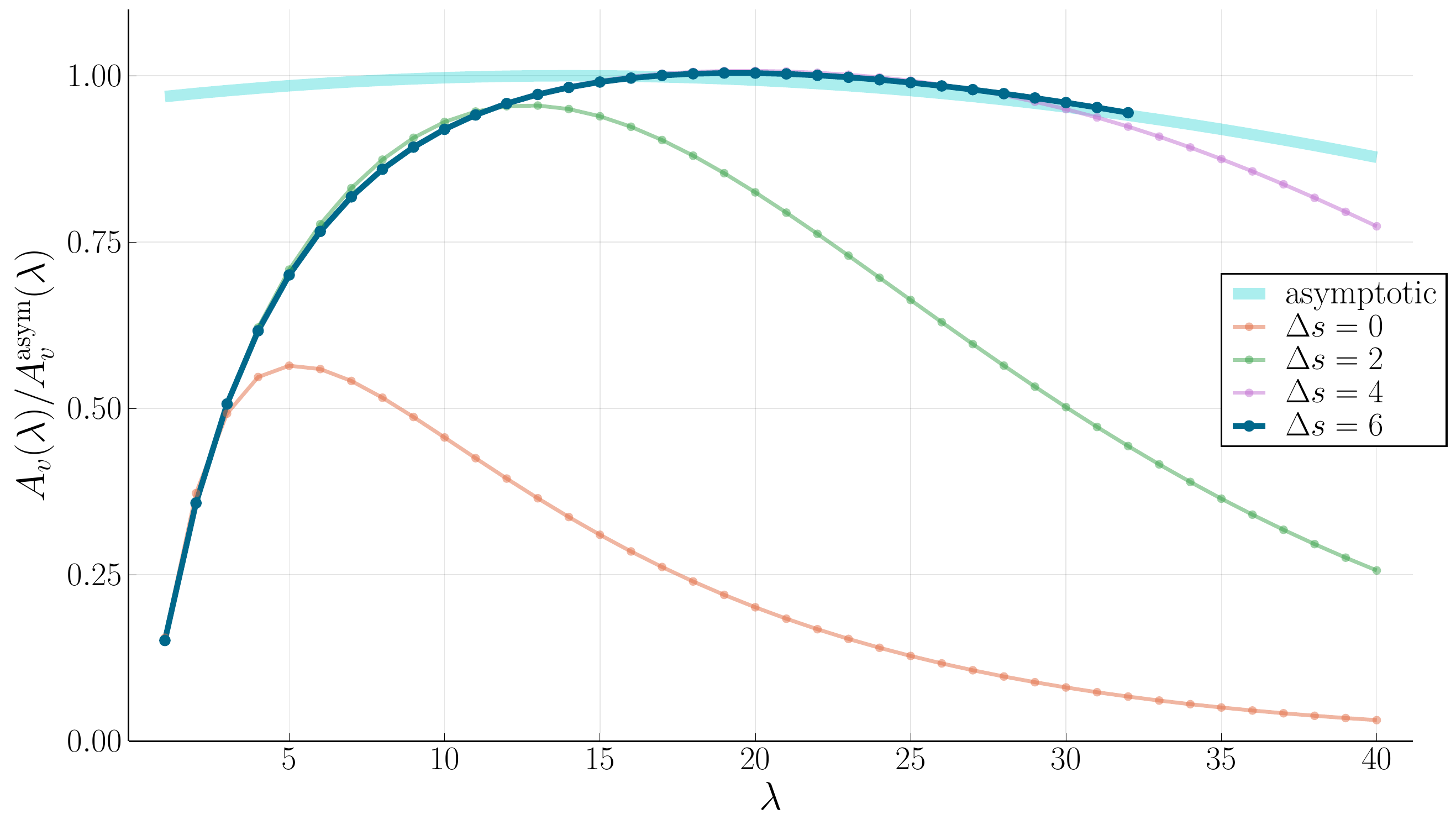}
    \end{subfigure}
    \vskip1cm
    \begin{subfigure}[b]{1.0\textwidth}
        \centering
        \includegraphics[width=\textwidth]{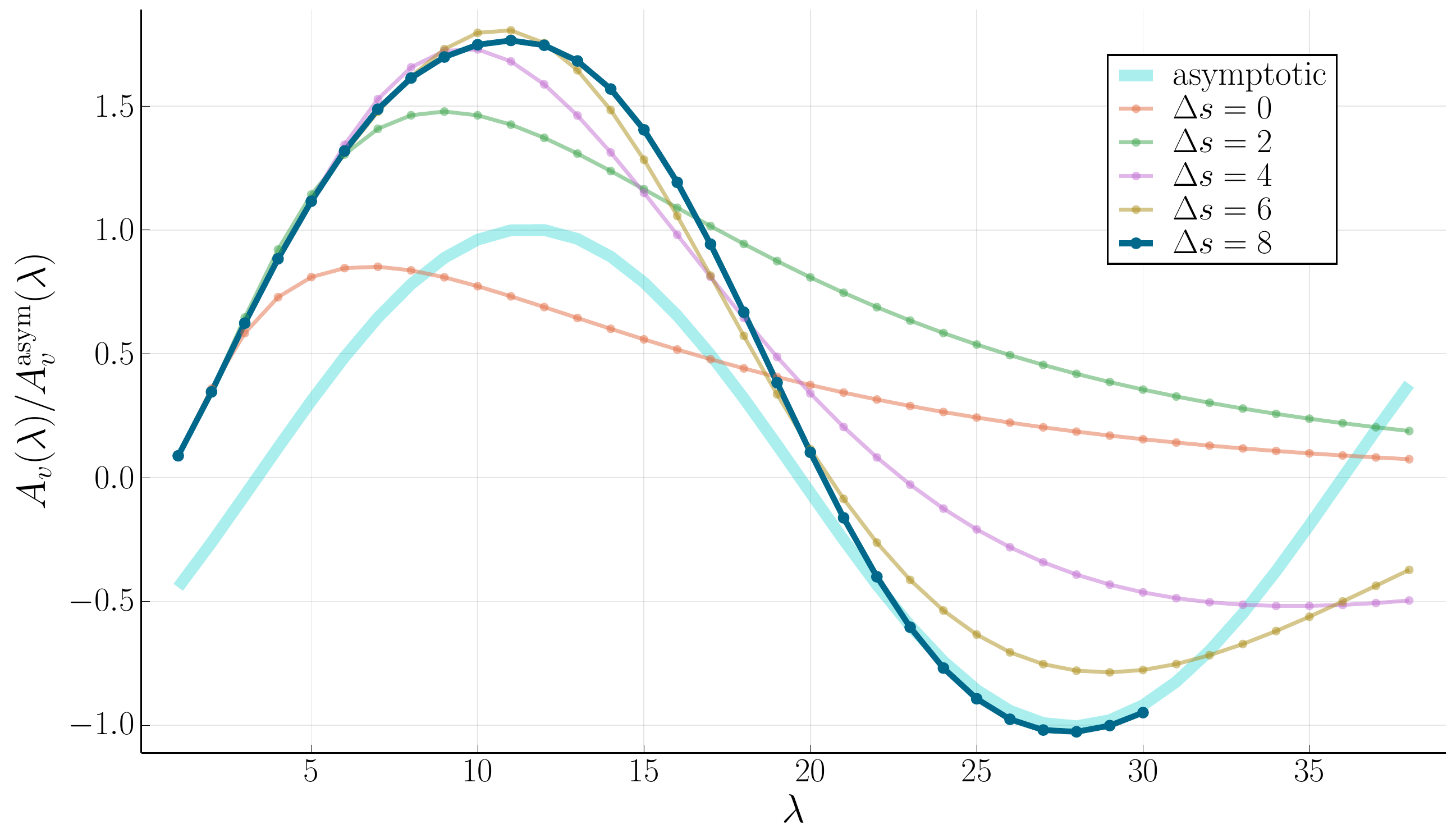}
    \end{subfigure}
    \vskip4mm
    \caption{Lorentzian asymptotics (real part, normalized). The computed curve with the highest number of shells is shown in dark blue. The exact asymptotic is shown as a thick light blue curve (the phase is chosen to approximately match the computed curve). \emph{Top:} Case $\gamma = 0.2$ up to $\Delta s = 6$ shells. The asymptotic amplitude is $\sim 2.95\cdot 10^{-13} \cos(0.27 - 0.02\lambda)$. \emph{Bottom:} Case $\gamma = 2.0$ up to $\Delta s = 8$ shells. The asymptotic amplitude is $\sim 1.17\cdot 10^{-17} \cos(2.22 - 0.19\lambda)$.}
    \label{fig:lor_asym}
\end{figure}
The plots are shown in Figure \ref{fig:lor_asym} for two different values of the Immirzi parameter $\gamma = 0.2$ and $\gamma = 2.0$. The two values of $\gamma$ show a markedly different behavior.
\begin{itemize}
    \item For the very slow oscillation of the case $\gamma = 0.2$ it is enough to set $\Delta s = 4$ to match the asymptotic value up to $\lambda \sim 30$. For higher spins, 4 shells are not enough to reach convergence to the exact amplitude and this results in the curve $\Delta s = 4$ to decay faster than the exact asymptotics. We can see that adding 2 more shells to reach $\Delta s = 6$ does not change the curve, hence convergence to the exact amplitude is reached for $\lambda \lesssim 30$.
    \item For the value $\gamma = 2.0$ one full period of oscillation is realized below $\lambda \sim 40$. Figure \ref{fig:lor_asym} shows that convergence is worse for a low number of shells in this case and 4 shells are barely enough to notice the oscillations of the amplitude. Increasing to 6 shells we start to see convergence to the first minima of the asymptotic amplitude. Ramping up to 8 shells we finally reach convergence to the asymptotic value for $\lambda \sim 30$. The point $\lambda = 30$ took about 18 hours of computation on $\sim 1500$ cpus. 
\end{itemize}

From the plots we see agreement between the numerical computation and the asymptotic amplitude if a sufficient number of shells is reached. We verify that the semiclassical regime is reached at $\lambda \sim 30$ for this particular configuration. This does not agree exactly with the estimate of leading-order corrections of \cite{hanNumericalComputationsNexttoleading2020}, whose authors find corrections of about $10\%$ to $20\%$ in cases similar to the present ones at $\lambda \sim 30$, however this might depend on the choice of the phase of the asymptotic amplitude. Interestingly, our results show that for this particular configuration the two limits $\lambda \to \infty$ and $\Delta s \to \infty$ are not independent, and the higher the value of $\lambda$, the more shells are needed to converge to the exact value of the amplitude. Unfortunately, this scaling is contrary to what one would hope, i.e.\ that a constant (and relatively low) number of shells would suffice to obtain a good approximation to the exact value of the amplitude even for high spins. However, a way out from this apparent cul-de-sac is to consider low values of $\gamma$, as in top plot of Figure \ref{fig:lor_asym}. For this particular value of $\gamma = 0.2$, a low number of shells well approximates the final amplitude till $\lambda \sim 30$.

\begin{jllisting}[caption={Lorentzian asymptotics.}, label={code:lorasym}]
    # vertex boundary spins
    js = [5 5 5 5 2 2 2 2 2 2];
        
    # spins along the strands of the edges 5,4,3,2,1
    jcs = [ [5 2 2 2], [2 5 2 2], [2 2 5 2], [2 2 2 5], [5 5 5 5] ];
    
    # angles for the boundary coherent states
    lor_angles = [ 
        [0.6154797086703873 ... -2.186276035465284],
        ...
    ];
    
    Lmax = 40
    
    # compute coherent states
    css = []
    for λ = 1:Lmax
        @time cs = [coherentstate_compute(λ .* jcs[i], lor_angles[i]) for i in 1:5]
        push!(css, cs)
    end
    
    # compute amplitudes
    shells = [ 0, 2, 4 ]
    ampls = [Array{ComplexF64}(undef, Lmax) for s in shells]
    
    for (i, s) in enumerate(shells)  
        @time for λ in 1:Lmax
            v = vertex_load(λ .* js, s)
            ampls[i][λ] = contract(v, css[λ]...)
        end 
    end
    
    # elaborate amplitudes (adjust phase, normalize etc)
    ...
    
    \end{jllisting}

\begin{figure}[!hbtp]
    \centering
    \begin{subfigure}[b]{0.49\textwidth}
        \centering
        \includegraphics[width=\textwidth]{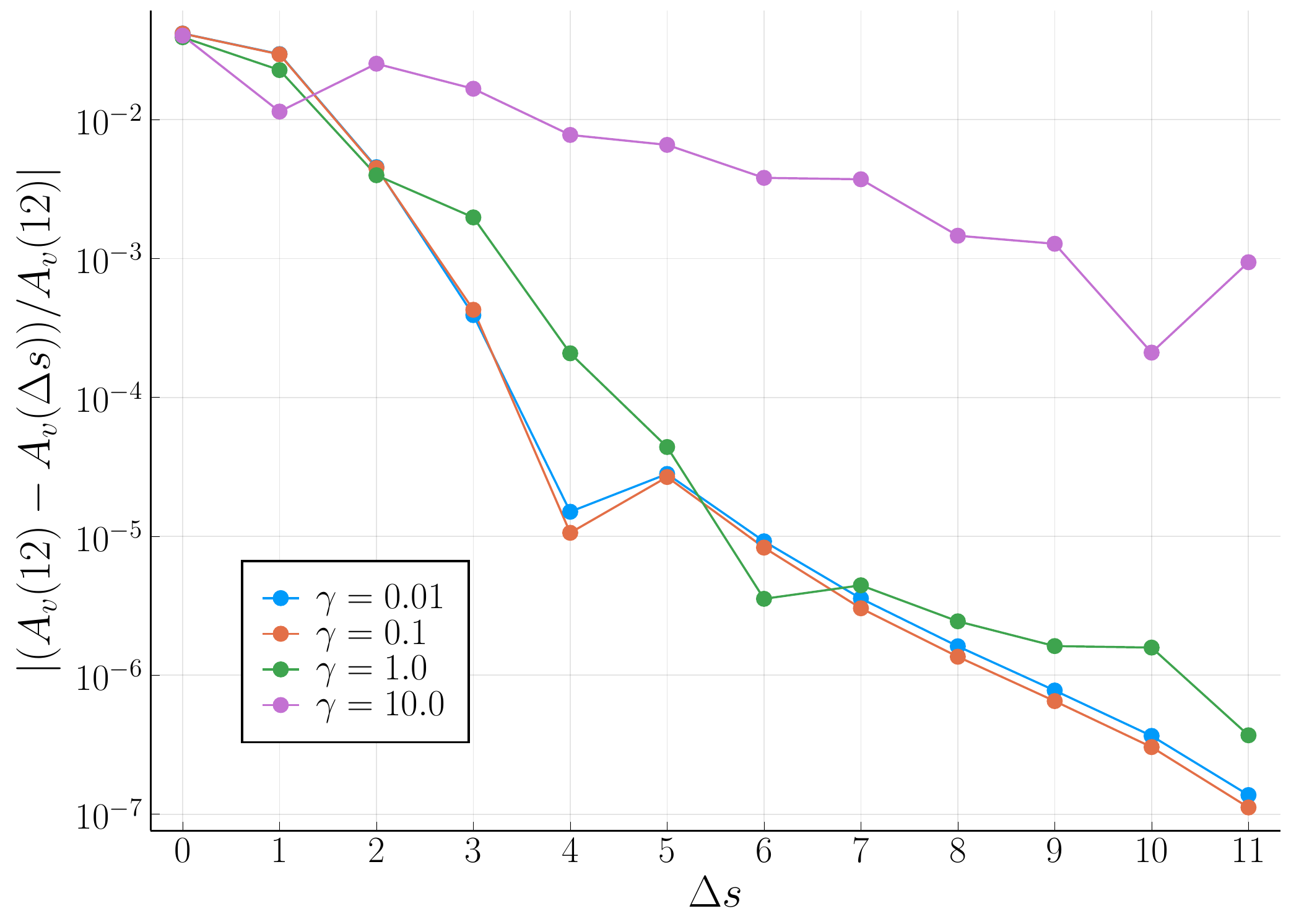}
    \end{subfigure}
    \hfill
    \begin{subfigure}[b]{0.49\textwidth}
        \centering
        \includegraphics[width=\textwidth]{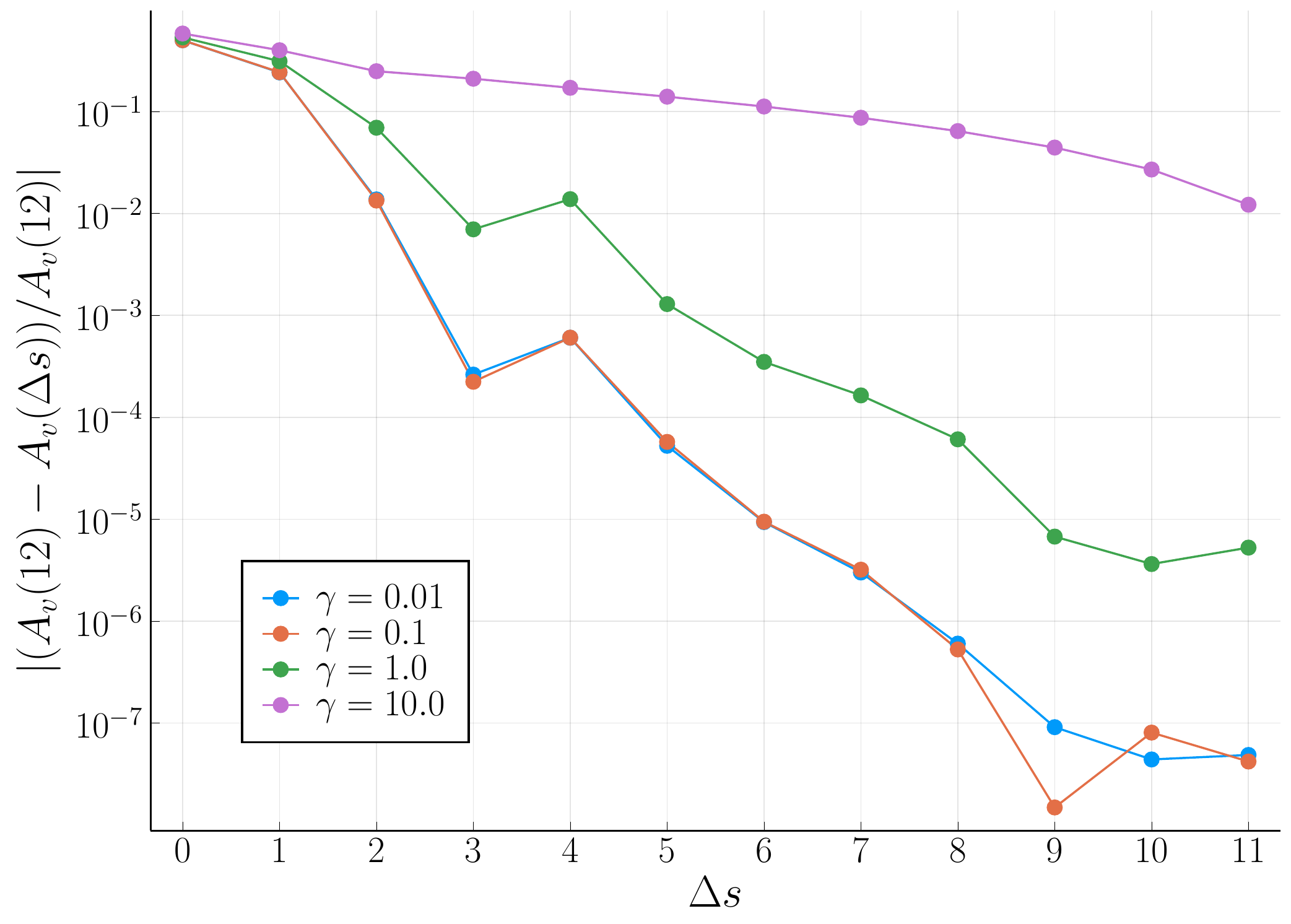}
    \end{subfigure}
       \caption{Log-plot of the convergence of the Lorentzian amplitude in the number of shells, measured as the relative error between the amplitude at $\Delta s =12$ and the amplitude with lower $\Delta s$. The convergence is shown for 4 different values of the Immirzi parameter $\gamma$. \emph{Left:} Case $\lambda = 2$. \emph{Right:} Case $\lambda = 10$.}
       \label{fig:lorasym-conv}
\end{figure}

To better quantify the previous remarks, we studied how the amplitude converges as a function of the number of shells, for different values of $\gamma$ and $\lambda$. In Figure \ref{fig:lorasym-conv} we plot the convergence of the amplitude for various values of $\gamma$, estimated as the relative error between the amplitude computed at $\Delta s = 0, 1, \ldots, 11$ shells versus the amplitude computed at $\Delta s = 12$ shells. The plots are for $\lambda = 2$ and $\lambda = 10$. The results show clearly that for values of $\gamma \ll 1$ convergence is fast in the number of shells, for both $\lambda$s. Conversely, for values of $\gamma$ of order 1 and higher a relatively high number of shells is required to reach convergence. This is especially evident in the case $\lambda = 10$. In light of Figure \ref{fig:lor_asym} we can easily deduce that the convergence further slows down at higher values of $\lambda$.

\vskip3mm

To summarize the results of this section, using \lib{} we have completed the task of testing numerically the asymptotic formula \eqref{eq:lor-asym}, initiated in \cite{donaNumericalStudyLorentzian2019}. We have highlighted the fact that, in the case of boundary data reconstructing to a Lorentzian 4-simplex, the ``shelled approximation'' introduced in \cite{spezialeBoostingWignerNjsymbols2017} works most effectively for values of $\gamma \ll 1$. It is interesting to compare these findings to the simpler case of an Euclidean 4-simplex \cite{donaSUGraphInvariants2018}, where instead a very low number of shells suffices to capture the correct asymptotics. The key technical difference between the two cases is likely to depend on the properties of the booster coefficients, which have been studied in \cite{donaAsymptoticsMathrmSLMathbbC2020} using the coherent representation. However, the exact mechanism by which these coefficients affect the present case is still not clear and would require further investigation.


\subsection{The $\dtre$ graph}

Although the above result about the single vertex asymptotics \cite{barrettAsymptoticAnalysisEnglePereiraRovelliLivine2009} is encouraging for connecting EPRL models to General Relativity in the semiclassical limit, the picture is less clear for the case of spin foams with many vertices. Indeed, the correct limit in which classical physics should be recovered is the double scaling in which the spins on the spin foam faces become large (low-energy limit) and the number of vertices increases (refinement). Whether or not it is possible to use the single vertex Regge asymptotics to study the refinement limit is, in our opinion, still an open question \cite{hanAsymptoticsSpinFoam2012,hanAsymptoticsSpinfoamAmplitude2013}. Various analysis in this direction pointed out what has been called the \emph{flatness problem} of EPRL models \cite{bonzomSpinFoamModels2009,conradySemiclassicalLimit4d2008,hellmannHolonomySpinFoam2013a,hanSpinfoamModelsLarge2014}. In its simplest form, the flatness problem states that the amplitudes that dominate the partition function \eqref{eq:partf-2} correspond to geometries which are flat in the semiclassical limit. More precisely, if we assume that the semiclassical analysis of the single vertex asymptotics holds also with many vertices and we identify certain quantities in the product of many vertices as the internal deficit angles of a Regge discretization, then the solutions to the dynamical equations, i.e.\ the dominant contributions to the path integral when taking the variation over the dynamical variables, correspond to vanishing deficit angles. This is believed to be a problem for EPRL models since apparently they cannot recover non-flat solutions of Einstein's equations (however, see \cite{hanSpinfoamModelsLarge2014} for a more careful interpretation that takes refinement into account).

An instance of the problem is given by a spin foam graph with 3 vertices connected to form a single internal face. We call this graph $\dtre$, following the literature. The $\dtre$ graph has been studied in various works \cite{magliaroCurvatureSpinfoams2011a,oliveiraEPRLFKAsymptotics2018,engleAddendumEPRLFK2021}, using the Euclidean theory. The results have been controversial, with the recent work by Engle et al.\ \cite{engleAddendumEPRLFK2021} concluding that the flatness problem appears as expected in the semiclassical analysis of the $\dtre$ graph once all the subtle mathematical details of the variation of the spin foam sum are taken into account. In \cite{donaNumericalAnalysisSpin2020} the authors proposed a method to attack the problem numerically. Using the previous version of the code \codek{sl2cfoam}, the authors attempted to compute the $\dtre$ amplitude numerically and study the saddle point structure of the sum over the internal face. They used the vertex of the topological BF theory as they claimed that the flatness problem should affect also the topological theory. They found that for certain values of the parameters which induced a curved geometry they could see the presence of a saddle point at the expected geometrical value. They inferred from this that the corresponding amplitude is not suppressed compared to flat amplitudes. However, their analysis was limited in the following aspects:
\begin{enumerate}
    \item the topological theory was studied in place of the physically relevant Lorentzian EPRL model;
    \item the range of boundary data was limited to only 2 distinct values of the parameter which regulates the curvature of the boundary geometry;
    \item the maximum value of the spins reached was satisfactory ($\sim 30$) only for the analysis of the flat case, which took more than two months of computational time.
\end{enumerate}
The last point requires particular care, as it is expected that the ``accidental constraint'' of the flatness problem which suppresses amplitudes manifests only at relatively large values of the spins involved \cite{engleAddendumEPRLFK2021}. Here we improve on every aspects of the previous numerical analysis using the new code \lib{}. We perform the same analysis of \cite{donaNumericalAnalysisSpin2020} using both the BF and Lorentzian EPRL vertex, for a larger selection of boundary data and a substantially larger value of the boundary spins. We are able to confirm the presence of the accidental constraint in both cases, using both an asymptotic and a saddle point analysis. Our analysis confirms that the flatness problem is present in the BF and EPRL spin foam models with the predicted accidental constraint. Whether this is a ``real'' problem or not is a question left for future works.

We refer the reader to \cite{donaNumericalAnalysisSpin2020} for all the details about the $\dtre$ graph, the exact formula for the amplitude in the BF case, the construction of the boundary geometry using Livine-Speziale coherent states and the numerical method to look for saddle points in the sum over the internal face. Here we only recall the formula that links the 4d dihedral angle $\alpha$ on the internal face with the geometrical area $x_g$ of the internal face:
\begin{equation}
    x_g = 3 \left( \frac{6 + \sin^2\alpha+6\sqrt{1-\sin^2\alpha}}{48+\sin^2\alpha} \right)^{\frac{1}{2}} \lambda
\end{equation}
where $\lambda$ is the common area of the boundary triangles. The function $x_g(\alpha)$ has period $\pi$ and is symmetric about $\alpha = \pi /2$. As a consequence, the deficit angle $\delta = 2\pi-3\alpha$ on the internal face has period $\pi$, which implies that the cases $\delta = 0$ and $\delta=\pi$ correspond to the same (flat) case, while the case $\delta = \pi/2$ is the most distant from the flat case, i.e.\ it is ``maximally curved''. For this reason we add to the analysis of \cite{donaNumericalAnalysisSpin2020} the case $\delta = 1.60$. By the same reasoning the cases $\delta = 2.47, 3.60$ reduce to the effective values $\delta \approx 0.67, 0.46$.\footnote{The previous considerations about the symmetries of $\delta$ as well as the case $\delta=1.60$ have been suggested to us by Hongguang Liu in a private communication.}

In Listing \ref{code:d3} we show the function that computes the coherent $\dtre$ amplitude given a vertex tensor \codek{v}, the spin \codek{x} on the internal face, the common boundary spin \codek{J} and the matrices for the angles of the boundary coherent states. Only one vertex tensor is required because, due to the symmetry of the problem, the three vertex tensors are all equal and have boundary spins $(x, J, J, \ldots)$. The vertex tensor can be computed on-the-fly (especially in the BF case) or again precomputed using MPI with the tool \codek{vertex-fulltensor} and then loaded into Julia.
\medskip
\begin{jllisting}[caption={$\Delta_3$ amplitude.}, label={code:d3}]
function d3(v, x, J, angles)::ComplexF64
   
    # compute coherent states
    css  = [ coherentstate_compute(J .* [1 1 1 1], angles[i])  for i in 1:9 ];
    
    # contract each vertex with boundary coherent states
    tA = contract(v, css[3], css[2], css[1])
    tB = contract(v, css[6], css[5], css[4])
    tC = contract(v, css[9], css[8], css[7])
    
    # now tA has indices (k2, k1)
    #     tB has indices (k1, k3)
    #     tC has indices (k3, k2)
    
    # intertwiners in the bulk
    rk, _ = intertwiner_range(J, J, J, x)
    ks = collect(rk[1]:rk[2])
    
    amx = 0.0
    
    # contract over internal intertwiners with phase
    for (ik1, k1) in enumerate(ks)
    for (ik2, k2) in enumerate(ks)
    for (ik3, k3) in enumerate(ks)
            
        amx += (-1)^(k1+k2+k3) * tA[ik2, ik1] * tB[ik1, ik3] * tC[ik3, ik2];       
                
    end
    end
    end
    
    (-1)^x * dim(x) * amx
    
end
\end{jllisting}

In the following we treat separately the BF and Lorentzian EPRL cases. We consider deficit angles $\delta_0 = 0$ (flat), $\delta_1 = 0.67$ (partially curved), $\delta_2 = 1.60$ (maximally curved). The corresponding geometrical internal areas are $x_{g0} = 1.34\lambda$, $x_{g1} = 1.26\lambda$, $x_{g2} = 1.14\lambda$.

\subsubsection*{BF vertex}

\begin{figure}
    \centering
    \begin{subfigure}[b]{0.49\textwidth}
        \centering
        \includegraphics[width=\textwidth]{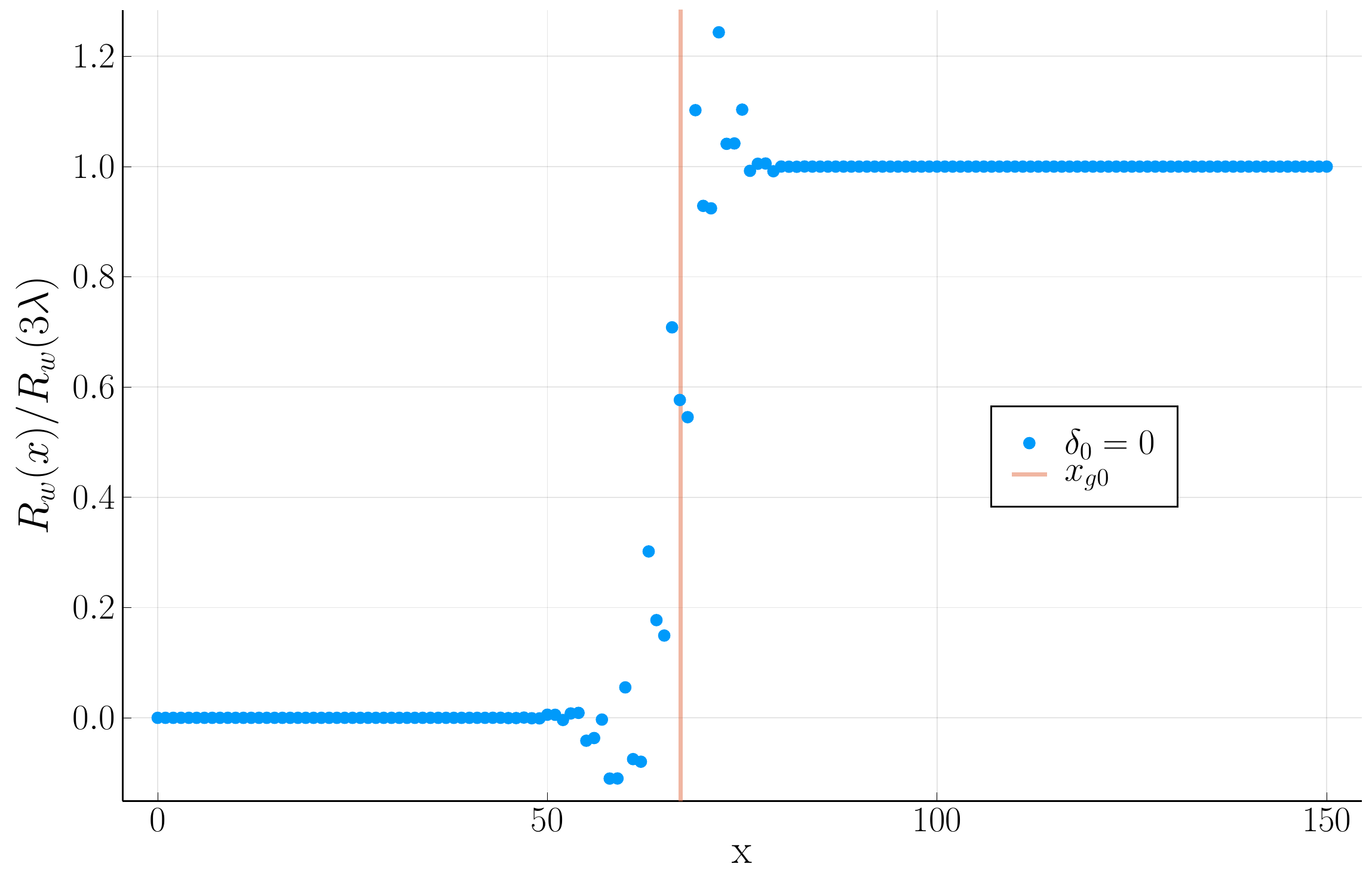}
    \end{subfigure}
    \hfill
    \begin{subfigure}[b]{0.49\textwidth}
        \centering
        \includegraphics[width=\textwidth]{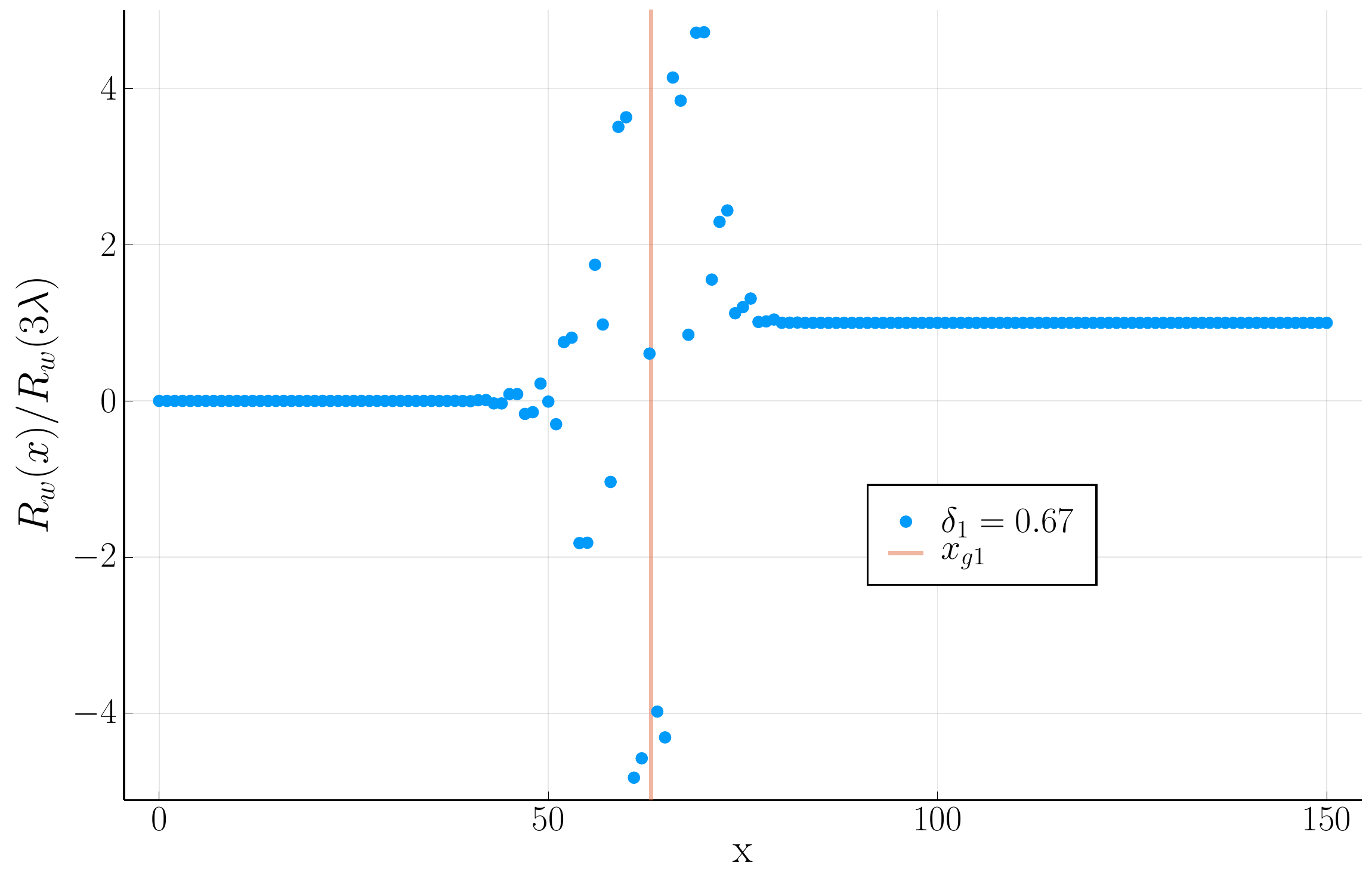}
    \end{subfigure}
    \hfill
    \vskip10pt
    \begin{subfigure}[b]{0.49\textwidth}
        \centering
        \includegraphics[width=\textwidth]{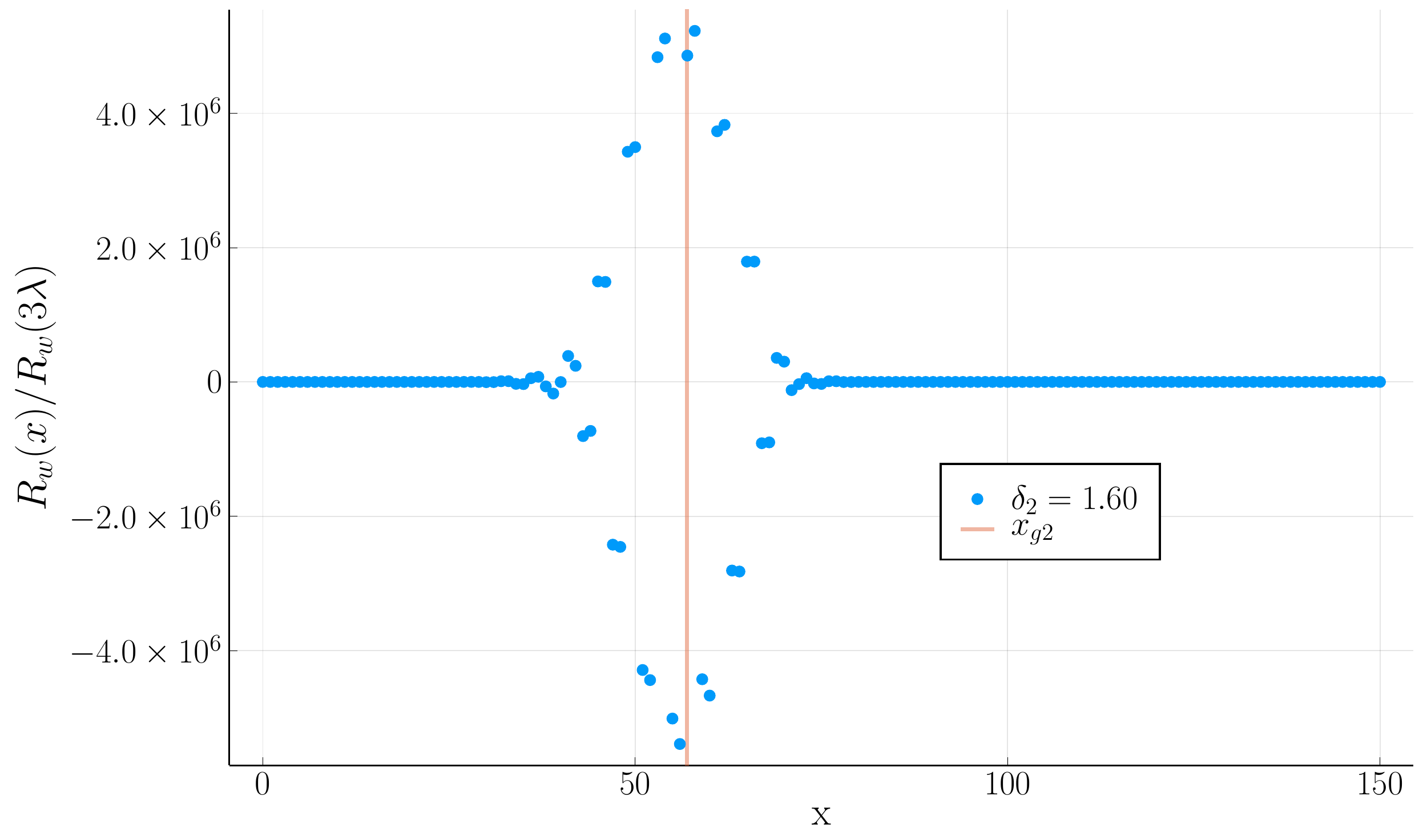}
    \end{subfigure}
       \caption{Numerical saddle point analysis of the sum over the internal spin $x$. The plots show the normalized partial sum $R_w(x)$ \cite{donaNumericalAnalysisSpin2020}. The presence of a saddle point contribution at the correct geometrical value $x_gi$ is evident in all cases. However in the case $\delta = 1.60$ there are large oscillations symmetric around 0 which almost exactly cancel, resulting in a final value which is suppressed compared to what would result only from the saddle point contribution. The same happens in the case $\delta = 0.67$ but the oscillations are much smaller, due to the very small suppression factor for this case.}
       \label{fig:d3_saddle}
\end{figure}

Using \lib{} we have been able to compute the $\dtre$ BF amplitude up to boundary spin $\lambda=50$. This last case took about one day of computation on a server with 112 cores. For comparison, the case $\lambda=30$ for various boundary configurations took about one hour on the same server, while the single flat configuration took more than 2 months of computation on multiple servers with the old code. We perform the same algorithmic saddle point analysis as in \cite{donaNumericalAnalysisSpin2020} for the case $\lambda=50$ using deficit angles $\delta_0, \delta_1, \delta_2$. The partial sums are shown in Figure \ref{fig:d3_saddle} along with the value of the geometrical internal area $x_{g}$. The presence of a saddle point in the sum along the internal spin $x$ is indicated by a ``jump'' in the amplitude at a location close to the geometrical value of $x$. The jump is evident in the flat case, with oscillations of order one at $x_{g0}$. The case $\delta_1$ shows again the presence of a saddle point, however the oscillations are considerably larger than the final value of the amplitude. Finally, the case $\delta_2$ has wild oscillations around $x_{g2}$ that completely mask the jump in the amplitude and the eventual presence of a saddle point. In this case the analysis is thus inconclusive.

We can explain the previous qualitative analysis by considering how the ``accidental constraint'' of the flatness problem is supposed to act \cite{hanSpinfoamModelsLarge2014}. We follow the recent analysis of Engle et al. \cite{engleAddendumEPRLFK2021}. Their conclusions about the Euclidean EPRL model can be adapted to the BF case by setting the Barbero-Immirzi constant to $\gamma=1$, in light of the similar asymptotic analysis of the two theories in the Euclidean sector and the following remarks. Their main result is the following formula:
\begin{equation}
    \label{eq:engle}
    Z(\lambda) \approx \lambda^k e^{i \lambda S_R^0} \sum_{k=-\infty}^{\infty} \exp \frac{-i\lambda}{4a} (4\pi k - \gamma\theta)^2 
\end{equation} 
for the partition function (the amplitude) of the $\dtre$ graph as a function of the scale $\lambda$. $S_R^0$ is the Regge action for the geometry corresponding to the chosen boundary data that induce internal deficit angle $\theta$. The ``accidental constraint'' has the explicit form
\begin{equation}
    \label{eq:acc-const}
    C_\gamma(\lambda, a, \theta) = \sum_{k=-\infty}^{\infty} \exp \frac{-i\lambda}{4a} (4\pi k - \gamma\theta)^2 
\end{equation} 
with $a$ a complex number which depends on different Hessians of the action of the quantum theory evaluated for the geometrical boundary data. The constraint seems to enforce the condition
\begin{equation}
    \theta \approx 0 \mod \frac{4\pi}{\gamma} 
\end{equation} 
with the symbol $\approx$ meaning in the asymptotic limit. Setting $\gamma = 1$ we recover the constraint expected in the BF case \cite{hellmannHolonomySpinFoam2013a}.

\begin{figure}
    \centering
    \includegraphics[width=.70\textwidth]{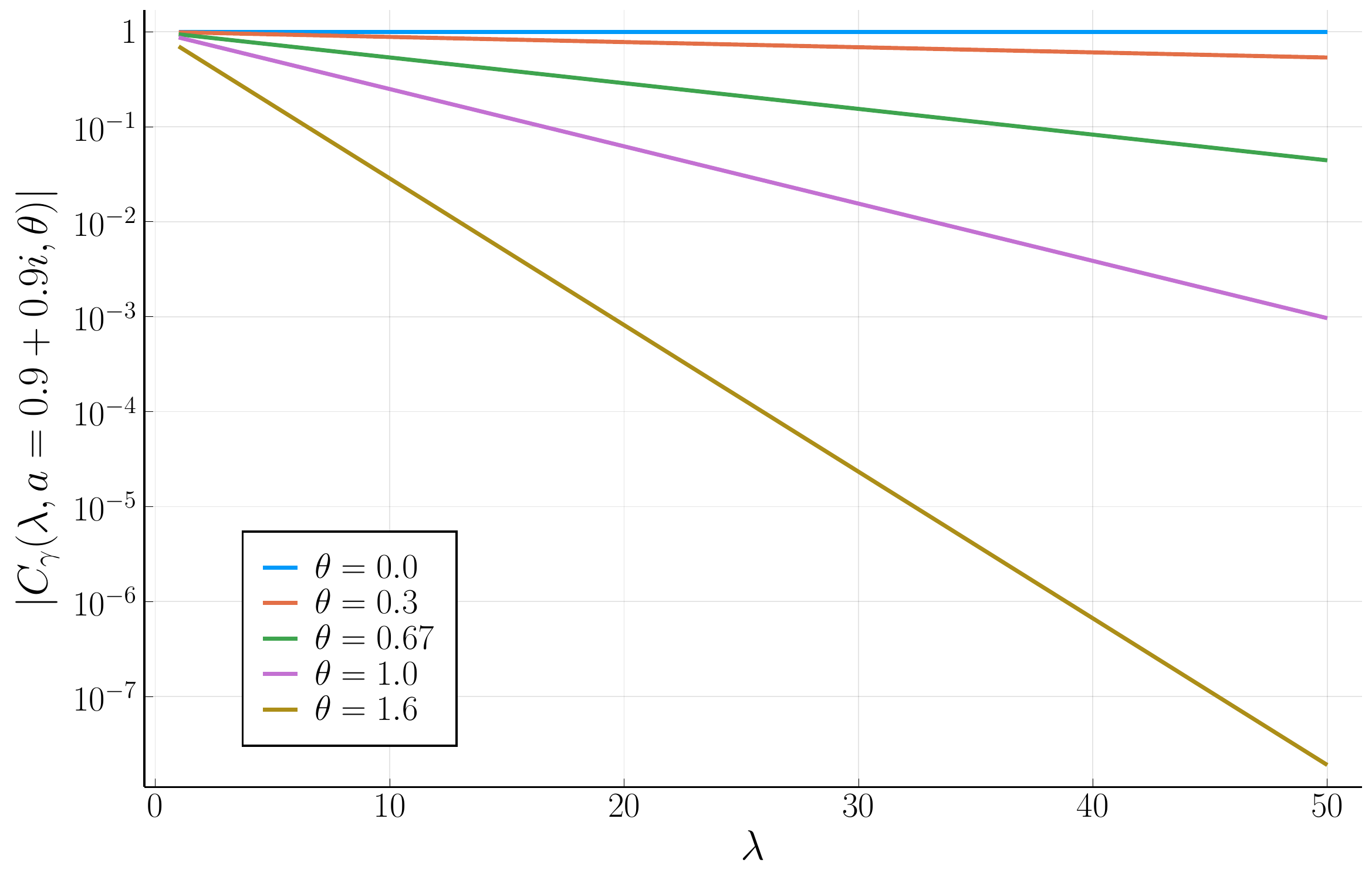}
    \caption{Log-plot of the accidental constraint $C_\gamma(\lambda, a, \theta)$ as estimated in \cite{engleAddendumEPRLFK2021}. The constraint is shown as a function of $\lambda$ for fixed $a = 0.9 + 0.9i$ and various values of $\theta$.}
    \label{fig:acc-const-exact}
\end{figure}

Numerically it is immediate to verify that $C_\gamma$ acts as a negative exponential in $\lambda$, with the magnitude of the suppression controlled by the deficit angle $\theta$. For small values of $\theta$ the decay is extremely slow, as we show in Figure \ref{fig:acc-const-exact} for the case $a = 0.9 + 0.9i$. The results of Figure \ref{fig:d3_saddle} show that the saddle point at $x_g$ is present and contributes the factor $e^{i \lambda S_R^0}$ in \eqref{eq:engle}, but the constraint $C_\gamma$ acts through the wild oscillations of the case $\delta_2$. These oscillations effectively cancel the the saddle point contribution and result in the slow suppression of the total amplitude.

\begin{figure}
    \centering
    \begin{subfigure}[b]{0.49\textwidth}
        \centering
        \includegraphics[width=\textwidth]{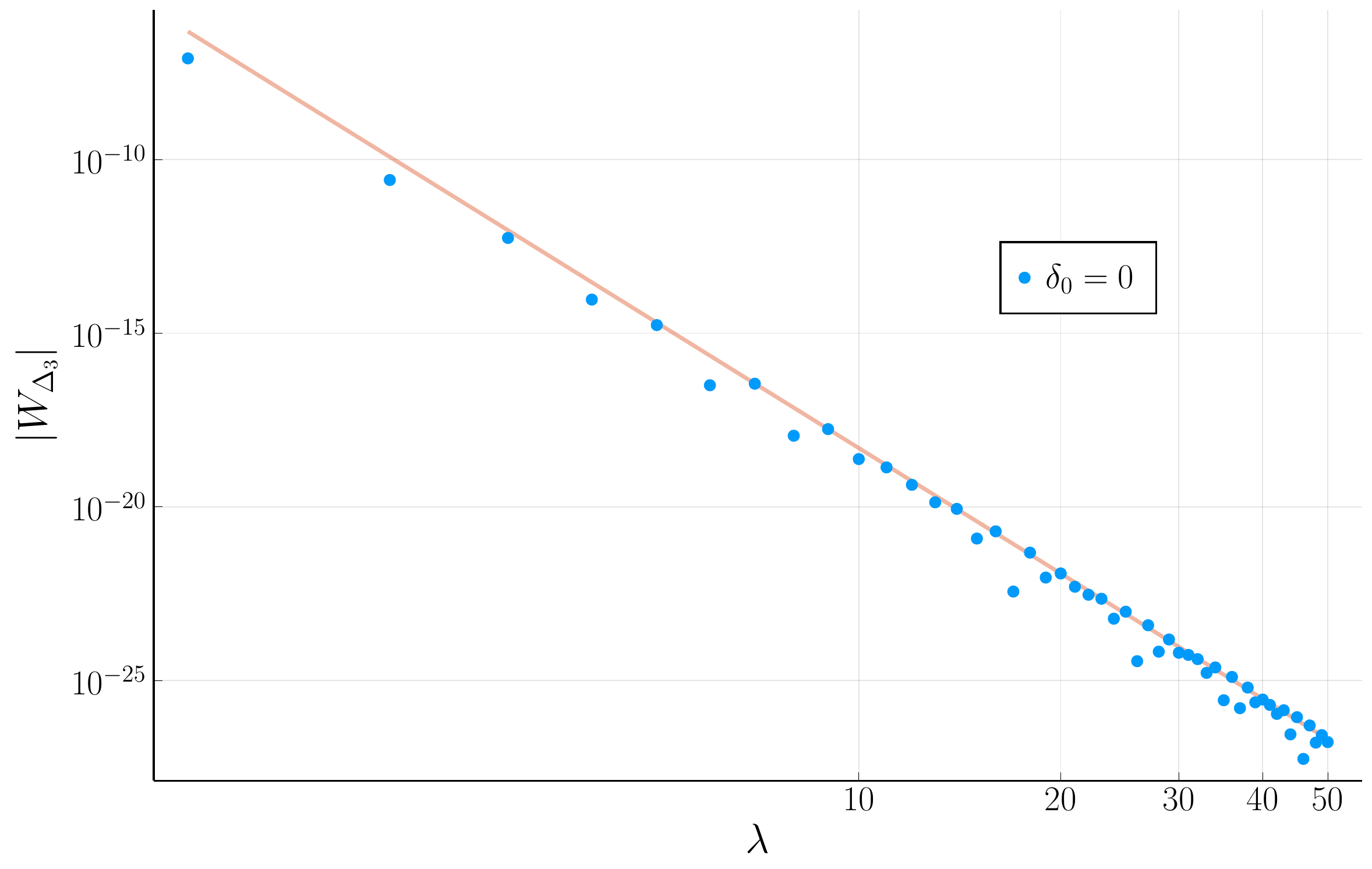}
    \end{subfigure}
    \hfill
    \begin{subfigure}[b]{0.49\textwidth}
        \centering
        \includegraphics[width=\textwidth]{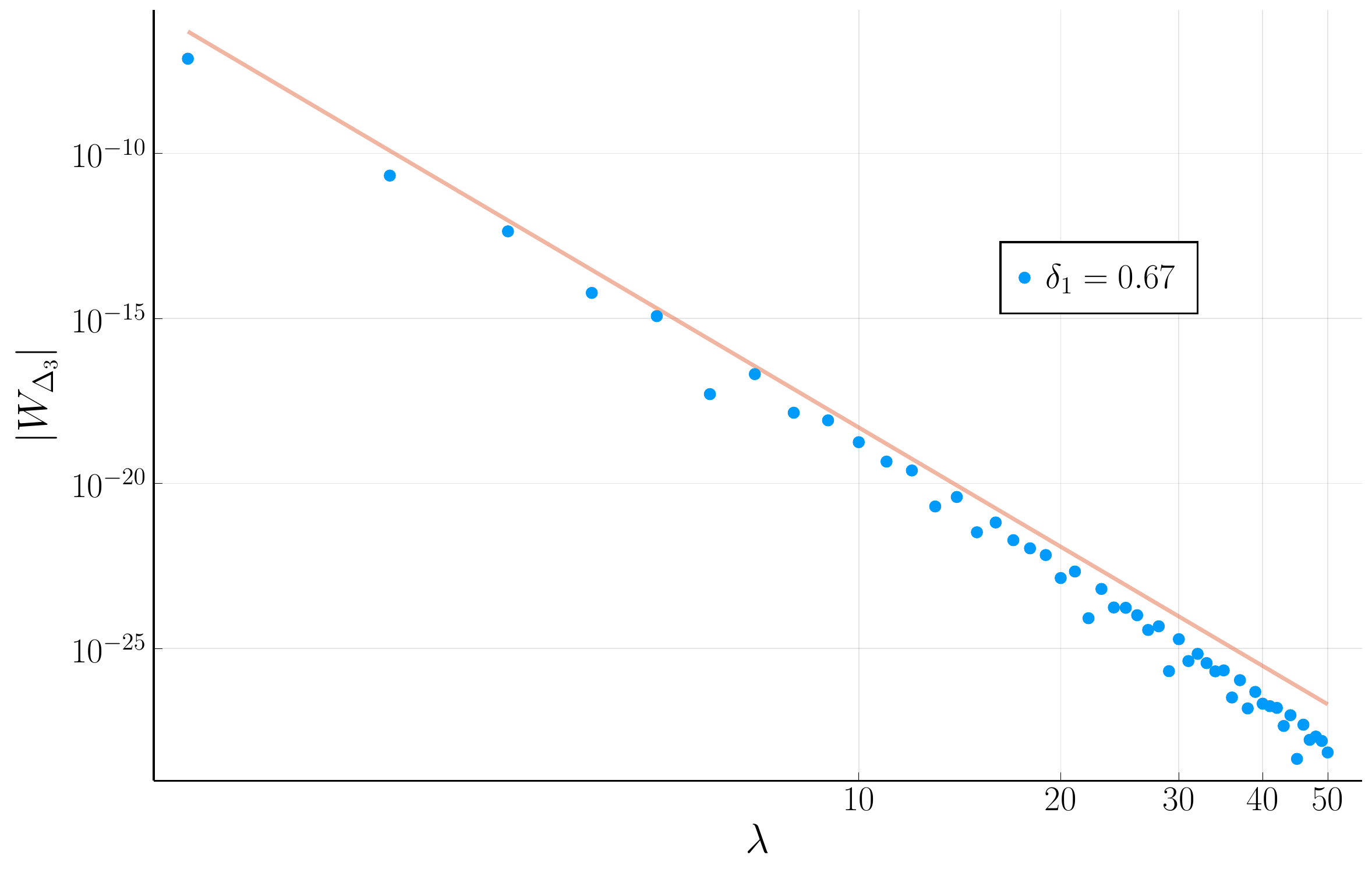}
    \end{subfigure}
    \hfill
    \vskip10pt
    \begin{subfigure}[b]{0.49\textwidth}
        \centering
        \includegraphics[width=\textwidth]{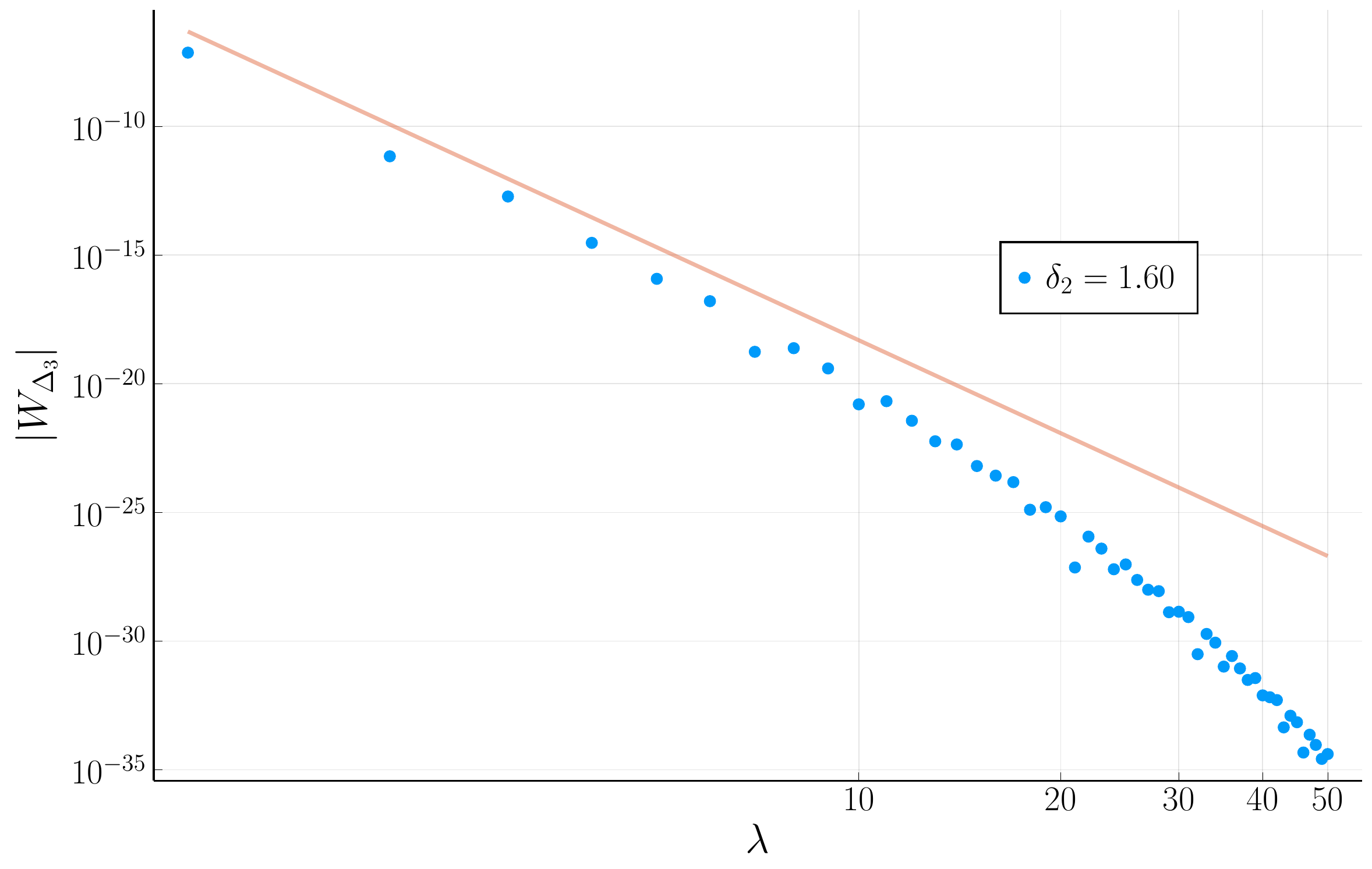}
    \end{subfigure}
       \caption{Numerical asymptotics of the $\dtre$ BF amplitude. The flat case ($\delta = 0$) follows the power law $\lambda^{-12}$. The case $\delta = 0.67$ shows a small suppression at high spins. In the case $\delta = 1.60$ the suppression is evident.}
       \label{fig:d3_asym}
\end{figure}

\begin{figure}[!hbtp]
    \centering
    \begin{subfigure}[b]{0.49\textwidth}
        \centering
        \includegraphics[width=\textwidth]{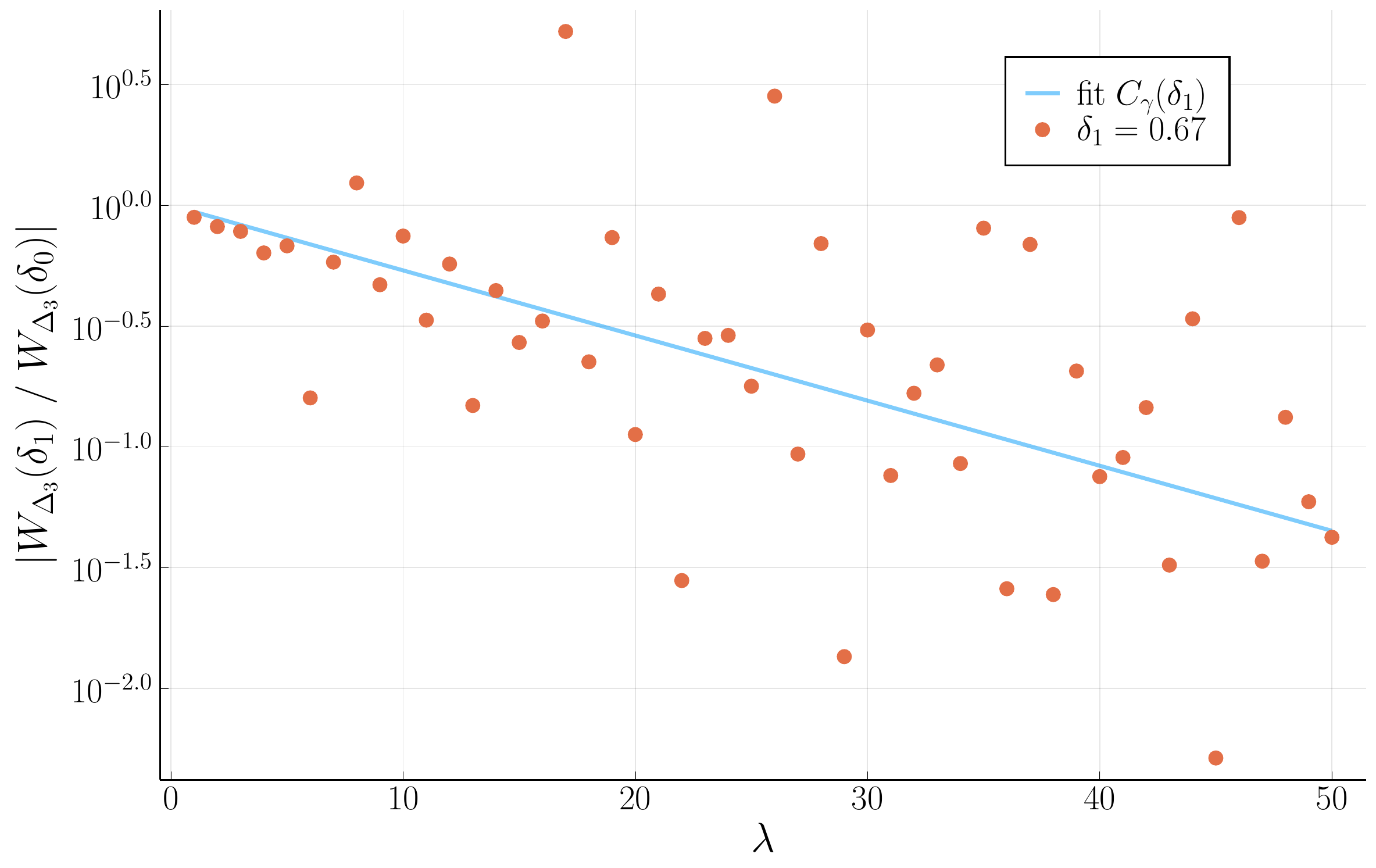}
    \end{subfigure}
    \hfill
    \begin{subfigure}[b]{0.49\textwidth}
        \centering
        \includegraphics[width=\textwidth]{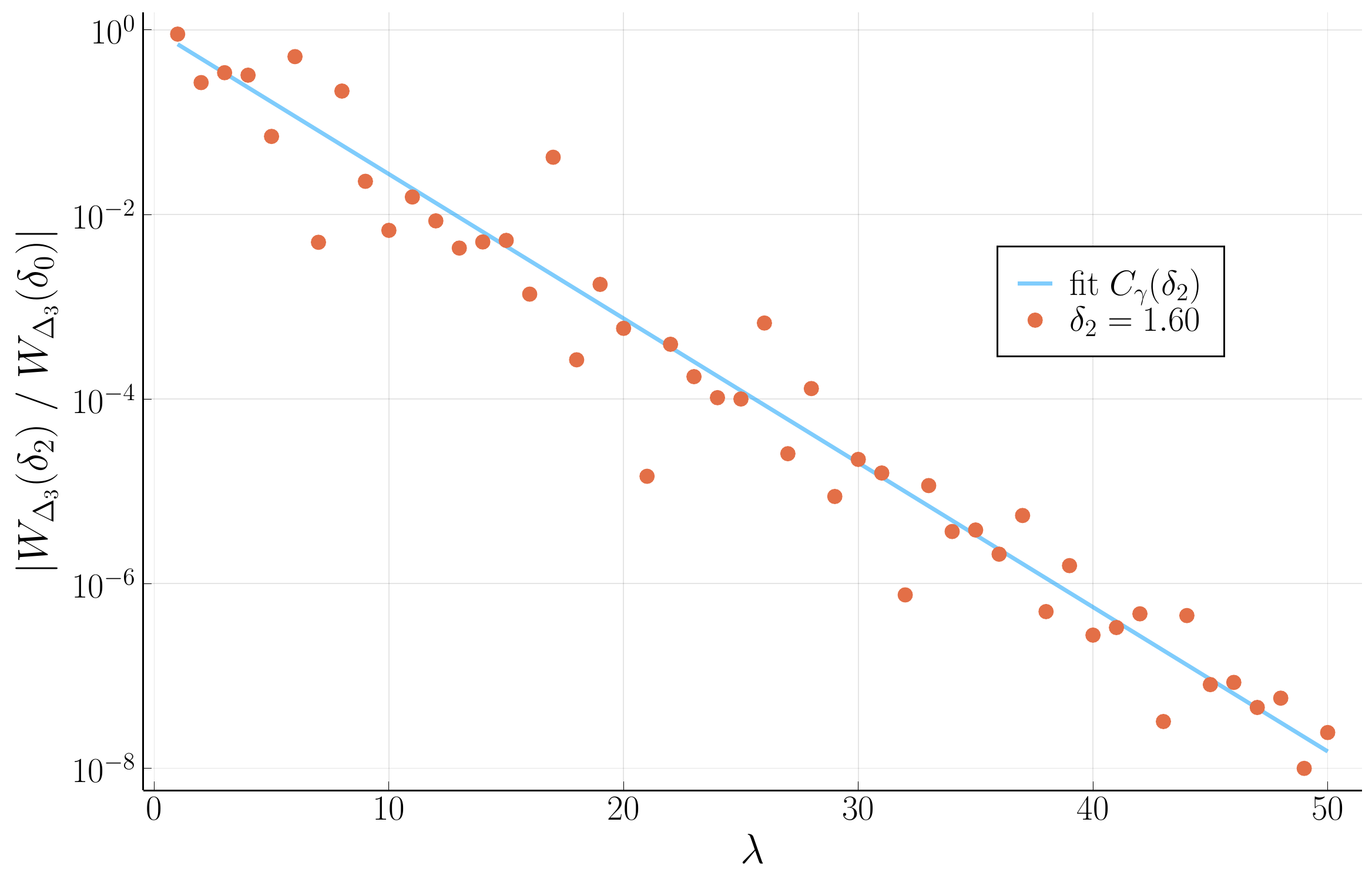}
    \end{subfigure}
       \caption{Numerical analysis of the accidental constraint $C_\gamma(\lambda, a, \delta)$ for BF theory. For $i=1,2$ the constraint $C_\gamma(\lambda, \cdot, \delta_i)$ is fitted to the computed suppression $W_{\dtre}(\lambda, \delta_i) \cdot \lambda^{12}$ with varying $a_i$. \emph{Left:} For the case $\delta = 0.67$ the fit is with $a_{\mathrm{fit}} = 0.90+0.99i$. \emph{Right:} For the case $\delta = 1.60$ the fit is with $a_{\mathrm{fit}} = 0.88+0.99i$.}
       \label{fig:d3_constraint}
\end{figure}

We can confirm the presence of the constraint $C_\gamma$ more precisely with an asymptotic analysis of the full $\dtre$ amplitude $W_\dtre(\lambda; \delta)$. We show in Figure \ref{fig:d3_asym} the asymptotic scaling in $\lambda$ of the amplitude, compared with power law $\lambda^{-12}$ which would result from a non-suppressed saddle point contribution. The suppression is evident in the case $\delta_2$. We can try to fit the constraint to the data to see if they are compatible with it. Figure \ref{fig:d3_constraint} shows the result of a fit of $W_\dtre(\lambda; \delta_i) / (\lambda^{-12})$ with $C_\gamma(\lambda, a_i, \delta)$ where the complex number $a_i$ is the parameter for the fit, for $i = 1,2$. The observed suppression is found to be compatible to the exponential decay of the predicted constraint $C_\gamma$ (compare also to Figure \ref{fig:acc-const-exact}) and both the estimated values of $a_i$ are close to the case considered in \cite{engleAddendumEPRLFK2021}. We conclude that the accidental constraint is present in the BF case and has the form \eqref{eq:acc-const} predicted in previous works \cite{hanSpinfoamModelsLarge2014,engleAddendumEPRLFK2021}.
\vskip8mm

\subsubsection*{EPRL vertex}

The analysis of the EPRL vertex shares many similarities with the BF case for the following reason. It is seen in the asymptotic analysis \cite{barrettAsymptoticAnalysisEnglePereiraRovelliLivine2009} that the phase of the Lorentzian EPRL model oscillates exactly as the BF model for geometries in the Euclidean sector. Namely, the dependence from the Immirzi parameter $\gamma$ is not in the frequency of the oscillations but only in the overall scaling through the Hessian of the action at the critical points. For this reason, in our case we should expect also the accidental constraint \eqref{eq:acc-const} not to depend on $\gamma$, the deficit angle of our $\dtre$ geometry being a standard (Euclidean) 4-dimensional angle. 
This point, apparently, has been until now overlooked in the literature since all the works so far have considered either the Euclidean EPRL model \cite{conradySemiclassicalLimit4d2008, bonzomSpinFoamModels2009,hellmannHolonomySpinFoam2013a} or the Lorentzian model in the Lorentzian sector \cite{hanSpinfoamModelsLarge2014}.

\begin{figure}[b]
    \centering
    \begin{subfigure}[b]{0.49\textwidth}
        \centering
        \includegraphics[width=\textwidth]{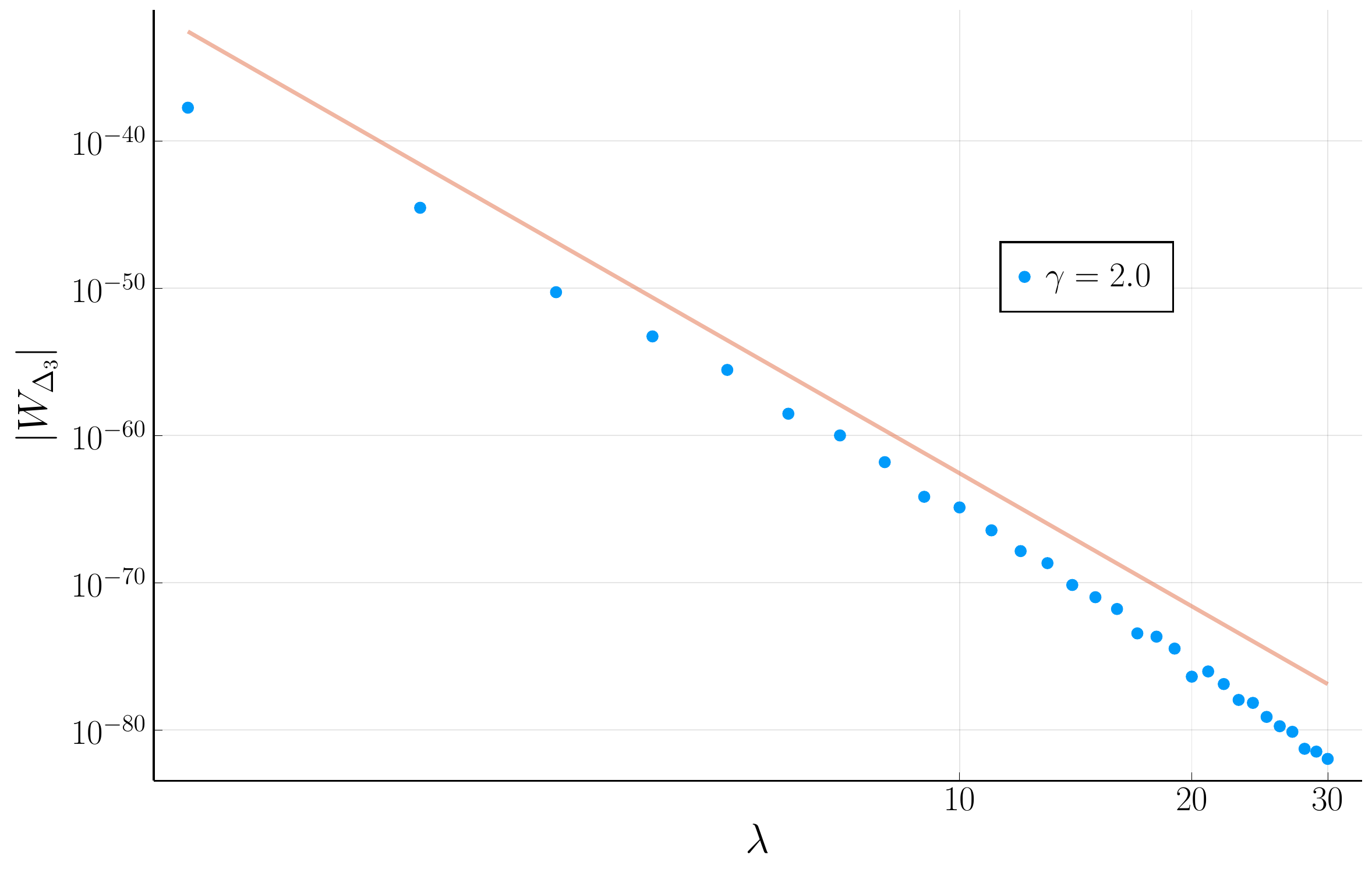}
    \end{subfigure}
    \hfill
    \begin{subfigure}[b]{0.49\textwidth}
        \centering
        \includegraphics[width=\textwidth]{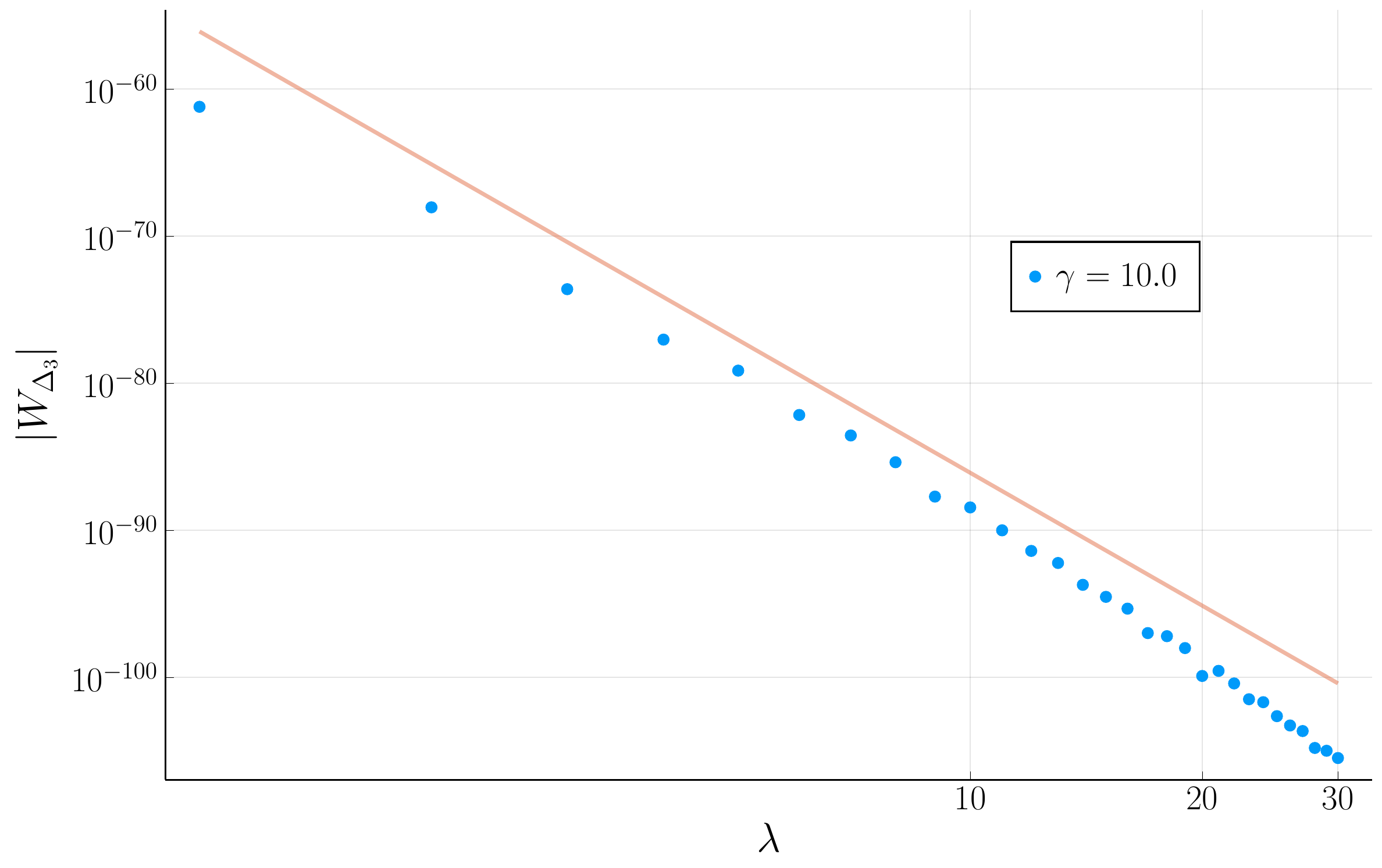}
    \end{subfigure}
       \caption{Numerical asymptotics of the $\dtre$ EPRL amplitude with $\Delta s = 2$. The internal deficit angle is $\delta = 1.60$. The straight line shows the unsuppressed asymptotic power law $\lambda^{-30}$. The two plots differ by the Immirzi parameter $\gamma$. \emph{Left:} Case $\gamma = 2.0$. \emph{Right:} Case $\gamma = 10.0$.}
       \label{fig:d3_eprl_asym}
\end{figure}

Numerically we can verify that the suppression of the BF case affects also the EPRL case. In Figure \ref{fig:d3_eprl_asym} we plot the asymptotic analysis of the EPRL $\dtre$ amplitude with $\Delta s = 2$ and $\lambda$ from 1 to 30, for two different values of the Immirzi parameter $\gamma = 2.0, 10.0$. We show only the maximally curved case $\delta = 1.60$ and compare it with the non-suppressed power law behavior $\sim \lambda^{-30}$. The two plots look exactly the same (except for the scale of the y-axis), reflecting the fact that neither the asymptotic oscillations nor the accidental suppression depend on $\gamma$. Isolating the suppression factor as in Figure \ref{fig:d3_constraint}, we can verify precisely that the degree of suppression is the same as the BF case and does not depend on $\gamma$. The plots are not shown since they are very similar to Figure \ref{fig:d3_constraint} --- although with a smaller range of $\lambda$.

\begin{figure}
    \centering
    \begin{subfigure}[b]{0.49\textwidth}
        \centering
        \includegraphics[width=\textwidth]{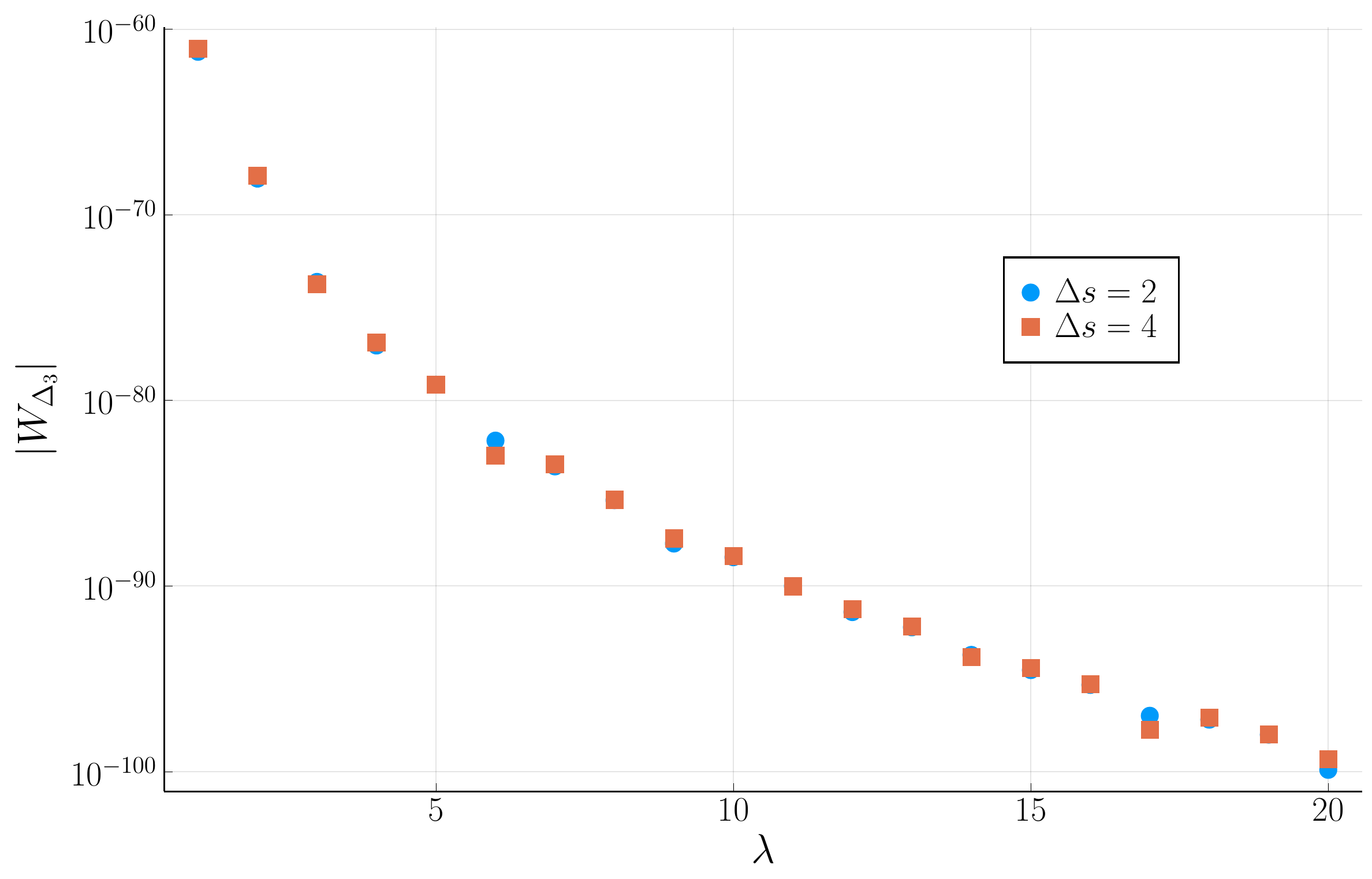}
    \end{subfigure}
    \hfill
    \begin{subfigure}[b]{0.49\textwidth}
        \centering
        \includegraphics[width=\textwidth]{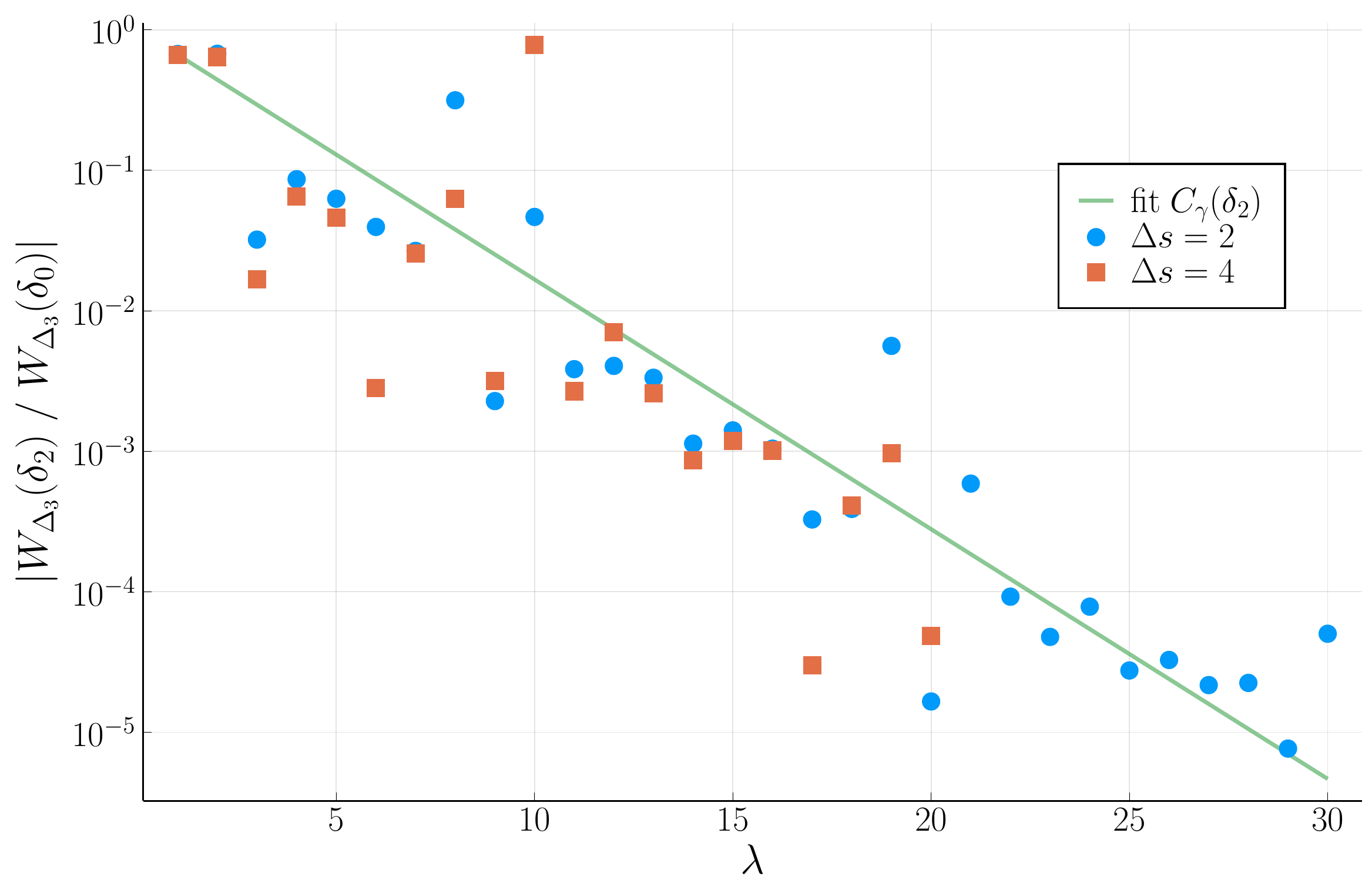}
    \end{subfigure}
       \caption{Comparison between the EPRL $\dtre$ asymtpotics at $\Delta s = 2$ and $4$, for $\gamma = 10.0$ and $\delta = 1.60$. \emph{Left:} Plot of the amplitudes. \emph{Right:} The constraint $C_\gamma(\lambda, \cdot, \delta_2)$ is fitted to the computed suppression $W_{\dtre}(\lambda, \delta_2) / W_{\dtre}(\lambda, \delta_0)$ with varying $a$. The fit is with $a_{\mathrm{fit}} = 0.78+0.82i$.}
       \label{fig:d3_eprl_shells}
\end{figure}

With the EPRL vertex, we also have to ensure that the cutoff in the number of shells $\Delta s$ does not affect the final result. We have two reasons to expect that a very low number of shells suffices in the present case. First, in the single vertex case it has been verified that a low number of shells captures immediately the asymptotic behavior of the Euclidean sector \cite{donaSUGraphInvariants2018}. Second, even supposing order-one fluctuations in the amplitude coming from the shells' approximation, the suppression due to the accidental constraint $C_\gamma$ is many orders of magnitude greater, especially in the case $\delta = 1.60$, and it is even stronger for large values of $\gamma$ (supposing a possible dependence on it). Therefore, we conclude that the suppression we see numerically is entirely due to the accidental constraint \eqref{eq:acc-const}. To reinforce this analysis, we computed the $\dtre$ amplitudes in the range $\lambda$ from 1 to 20 also at $\Delta s = 4$. We show in Figure \ref{fig:d3_eprl_shells} that the higher number of shells results only in small fluctuations from the $\Delta s = 2$ case and, most importantly, the suppression factor is the same for the overlapping range of $\lambda$s.

\vskip8mm
%


\section{Summary and next steps}
\label{sec:conclusions}

Calculations with Lorentzian EPRL spin foam amplitudes can be a hard feat. Analytically, not much is known outside of the large spin regime where saddle point approximations can be used, and even in this case dealing with all the complexities and subtleties of the integral expressions for most but the simplest graphs can be overwhelming. While the Euclidean EPRL model is more manageable, the Lorentzian model has been studied analytically only in extremely simple configurations \cite{barrettAsymptoticAnalysisEnglePereiraRovelliLivine2009, bianchiLQGPropagatorNew2009} or under drastic approximations \cite{rielloSelfenergyLorentzianEnglePereiraRovelliLivine2013, christodoulouPlanckStarTunneling2016}. Further progress using these methods seems, at the moment, unlikely at best (however, see \cite{donaWickRotationEPRL2021a} for a recent work that relates the Euclidean and Lorentzian models through a ``Wick rotation'' of the Immirzi parameter).

Numerical computations provide a novel and alternative way to explore all the features of spin foam models. Numerical codes have their own limitations, but these are usually of different nature than the common analytical approximations, hence they can offer a different perspective on the subject and can complement the analytical works in various respects. Numerical simulations have been applied, for example, to study the renormalization flow of the Euclidean EPRL model \cite{bahrHypercuboidalRenormalizationSpin2017} or, recently, to the problem of finding universal features of spin foam models using effective actions in both signatures \cite{asanteEffectiveSpinFoam2020,asanteEffectiveSpinFoam2021}. In the case of the Lorentzian EPRL model, which ideally should be employed in simulations due to its physical relevance, there have been already a few notable numerical works \cite{donaSUGraphInvariants2018,donaNumericalStudyLorentzian2019,gozziniPrimordialFluctuationsQuantum2021,donaNumericalAnalysisSpin2020}, using the library \codek{sl2cfoam} \cite{donaNumericalMethodsEPRL2018} or Montecarlo integration \cite{hanSpinfoamLefschetzThimble2021}. Regarding \codek{sl2cfoam}, unfortunately, the lack of optimization and the difficulty of using the library has prevented further work from different authors and on more complex problems.

In this work we presented the library \lib{}, a completely rewritten version of \codek{sl2cfoam}. The new code solves most of the shortcomings of the previous version: it is more stable and precise, it increases the performance by many orders of magnitude, it can scale over a large number of CPUs or on the GPU and it has a user-friendly scripting interface. We have shown how these goals have been met by using techniques and ideas ranging from High Performance Computing to tensor networks. We provided many examples of how to use the Julia interface for quick and interactive visualization of non-trivial computations within the Lorentzian EPRL model. Finally, to show more involved applications, we have presented the numerical study of two problems coming from spin foam literature. First, we have completed the numerical test of the Lorentzian asymptotics of the single vertex, which was initiated in \cite{donaNumericalStudyLorentzian2019}. We have highlighted how the convergence properties of the ``sum over shells'' of \cite{spezialeBoostingWignerNjsymbols2017} depend on the value of the Immirzi parameter, with values $\gamma \ll 1$ ensuring better convergence. Second, we have computed the transition amplitude of a graph made by 3 vertices and we have shown that the ``flatness problem'' of spin foam model emerges both in the BF and Lorentzian EPRL case. This settles a question that is still actively debated \cite{bianchiPanelVertex2020}.

The possible applications of \lib{} are many, and some are already being addressed. Here we only list a few of them.

Some applications are extensions of the present work. First, it is necessary to better understand the relation between the number of shells employed, the value of the Immirzi parameter, the average value of the boundary spins and the signature of the reconstructed geometry. Second, for the $\dtre$ graph, it would be useful to verify that and how the accidental constraint acts on boundary data that reconstruct a curved Lorentzian geometry (i.e.\ with non-zero Lorentzian deficit angle). A Lorentzian curved simplicial manifold has been recently described in \cite{asanteEffectiveSpinFoam2021}, although the value of the boundary spins seem too large to be studied outside of effective spin foams, where closed exact formulae can be found.

Going beyond the $\dtre$ graph and the flatness problem, the next step would be to study a triangulation with non-trivial Regge dynamics, similarly to what has been done by Asante et al. \cite{asanteEffectiveSpinFoam2020,asanteEffectiveSpinFoam2021} using the (computationally much simpler) effective spin foam models. Then, either by computing bulk observables or by searching numerically for saddle points \cite{donaSearchingClassicalGeometries2020} it would be possible in principle to understand whether the semiclassical limit of the EPRL dynamics matches that of Regge calculus in the general setting. This in turn would clarify if EPRL spin foams can recover general relativity in the semiclassical limit. Some preliminary studies about bulk observables are in progress \cite{gozziniHighPerformanceCode2021}.

While it is known that spin foams graphs with bubbles (i.e.\ internal dynamical faces that form a topological sphere) are infrared divergent in the standard, non-quantum-deformed theory, not much is clear about the degree of the divergence. The only works that tackled this question have provided a logarithmic lower bound \cite{rielloSelfenergyLorentzianEnglePereiraRovelliLivine2013} or a large polynomial upper bound \cite{donaInfraredDivergencesEPRLFK2018} for the self-energy of the ``melon'' graph. This is a graph with 6 faces and 4 intertwiners in the bulk, and 4 faces and 2 intertwiners on the boundary. The numerical simulation of this spin foam is in progress \cite{frisoniNumericalAnalysisEPRL2021} to refine the known bounds or even find the exact degree of divergence. Related to this, an intriguing possibility would be to see if the divergences can be eventually cured by the running of the Immirzi parameter, similarly to what has been studied in \cite{benedettiPerturbativeQuantumGravity2011} within the classical theory.

Another obvious application of \lib{} would be in the study of the renormalization flow of the Lorentzian theory, extending what has already been done in the Euclidean theory \cite{bahrHypercuboidalRenormalizationSpin2017}. A priori, the tensorial vertices that \lib{} manipulates could be adapted to apply proper tensor networks methods such as Tensor Network Renormalization \cite{evenblyTensorNetworkRenormalization2015} to study large graphs with low quantum numbers. An implementation of these techniques would shed much light on the coarse graining of the theory \cite{steinhausCoarseGrainingSpin2020}.

The possible applications are not limited to purely theoretical questions. Spin foam models of physical phenomena have already been proposed in the literature, and numerical simulations would help to study their possible physical implications. An extension to many vertices of the ``no-boundary'' cosmological model proposed in \cite{gozziniPrimordialFluctuationsQuantum2021} is currently being worked on. Preliminary investigations have also started for applying numerical techniques to the black hole-to-white hole transition, where an explicit formulation suitable to our code has appeared recently \cite{soltaniEndBlackHole2021}.

We also would like to remark that \lib{} can be used for educational purposes. The formalism and language of LQG take much time to be grasped by newcomers to the community, and this is especially true for spin foam models and the construction of the EPRL vertex. The Julia interface of \lib{} can be used on a laptop to visualize interactively what is a vertex amplitude, how does it scale, how the employed approximations work, how to combine multiple vertices, how to interpret geometrically the coherent amplitudes and so on. This would surely benefit students for easy visualization of otherwise abstract quantities. Moreover, we believe that also researchers in the field might find the library useful for rapid prototyping and testing of simple ideas and toy models, before embarking on more complicated calculations.

We conclude with a non-exhaustive list of possible improvements to \lib{}. Firstly, it could be useful to write a version of the library for the Euclidean EPRL model, which is computationally much simpler: booster coefficients are products of $9j$-symbols and the parameter $\Delta s$ has a natural upper bound. The Euclidean version could be used for quick testing of new ideas and models or can be used to infer results about the Lorentzian sector through analytic continuation \cite{donaWickRotationEPRL2021a}. For the present code, the integration grid for the booster function could be improved to adapt better to the values of the spins and the Immirzi parameter $\gamma$, saving time and increasing precision. Machine learning techniques could be used in this respect to guess the best range of integration given the spins $j_i, l_i$ and $\gamma$. Currently, the focus of the parallelization is on computing a few vertex tensors with a large number of shells. Complementary, we could develop routines for parallelizing many vertex tensors with a relatively low number of shells. These are necessary improvements for dealing with large graphs with many internal faces. Another step forward would come from finding a way to leverage the GPU in the computation of the vertex tensors, and not only in the contraction phase. Finally, better tools and a detailed documentation would certainly benefit the future users of the library.

\section{Acknowledgments}

The Centre de Calcul Intensif d'Aix-Marseille is acknowledged for granting access to its high-performance computing resources. We thank Pietro Don\`{a}, Pietropaolo Frisoni, Carlo Rovelli and Simone Speziale for many useful comments about the code and its first applications.






\printbibliography

\end{document}